\documentclass[pdflatex,sn-mathphys-num]{sn-jnl}

\usepackage{graphicx}%
\usepackage{multirow}%
\usepackage{amsmath,amssymb,amsfonts}%
\usepackage{amsthm}%
\usepackage{mathrsfs}%
\usepackage[title]{appendix}%
\usepackage[dvipsnames]{xcolor}
\usepackage{textcomp}%
\usepackage{manyfoot}%
\usepackage{booktabs}%
\usepackage{algorithm}%
\usepackage{algorithmicx}%
\usepackage{algpseudocode}%
\usepackage{listings}%
\usepackage{subcaption}%
\usepackage{multirow} %

\usepackage{marginnote}
\usepackage{tikz}
\usepackage[absolute,overlay]{textpos}
\usepackage{cleveref}

\theoremstyle{thmstyleone}%
\theoremstyle{thmstyletwo}%

\theoremstyle{thmstylethree}%

\begin{document}

\title[Article Title]{A Cyclic Layerwise QAOA Training}


\author[1]{\fnm{Enhyeok} \sur{Jang}}\email{enhyeok.jang@yonsei.ac.kr}

\author[2]{\fnm{Zihan} \sur{Chen}}\email{zihan.chen.cs@rutgers.edu}

\author[3]{\fnm{Dongho} \sur{Ha}}\email{dongho9601@gmail.com}

\author[1]{\fnm{Seungwoo} \sur{Choi}}\email{seungwoo.choi@yonsei.ac.kr}

\author[1]{\fnm{Yongju} \sur{Lee}}\email{yongju.lee@yonsei.ac.kr}

\author[1]{\fnm{Jaewon} \sur{Kwon}}\email{jaewon.kwon@yonsei.ac.kr}

\author[2]{\fnm{Eddy Z.} \sur{Zhang}}\email{eddy.zhengzhang@gmail.com}

\author[2]{\fnm{Yipeng} \sur{Huang}}\email{yipeng.huang@rutgers.edu}

\author*[1]{\fnm{Won Woo} \sur{Ro}}\email{wro@yonsei.ac.kr}

\affil*[1]{\orgdiv{School of Electrical and Electronic Engineering}, \orgname{Yonsei University}, \orgaddress{\city{Seoul}, \country{Korea}}}

\affil[2]{\orgdiv{Department of Computer Science}, \orgname{Rutgers University}, \orgaddress{\city{Piscataway}, \state{NJ}, \country{United States}}}

\affil[3]{\orgdiv{Unaffiliated}, \city{Seoul}, \country{Korea}}

\abstract{
The quantum approximate optimization algorithm (QAOA) is a hybrid quantum-classical algorithm for solving combinatorial optimization problems.
Multi-angle QAOA (MA-QAOA), which assigns independent parameters to each Hamiltonian operator term, achieves superior approximation performance even with fewer layers than standard QAOA.
Unfortunately, this increased expressibility can raise the classical computational cost due to a greater number of parameters.
The recently proposed Layerwise MA-QAOA (LMA-QAOA) reduces this overhead by training one layer at a time, but it may suffer from obtaining the precise solution due to the previously fixed parameters.
This work addresses two questions for efficient MA-QAOA training: 
(i) What is the optimal granularity for parameter updates per epoch, and 
(ii) How can we get precise final cost function results while only partially updating the parameters per epoch?
Despite the benefit of reducing the parameters that update per epoch can reduce the classical computation overhead, too fine or coarse a granularity of Hamiltonian update can degrade the MA-QAOA training efficiency.
We find that optimizing one complete layer per epoch is an efficient granularity.
Moreover, selectively retraining each layer by tracking gradient variations can achieve a final cost function equivalent to the standard MA-QAOA while lowering the parameter update overhead.
Based on these insights, we propose Orbit-QAOA, which cyclically revisits layers and selectively freezes stabilized parameters.
Across diverse graph benchmarks, Orbit-QAOA reduces training steps by up to 81.8\%, reduces approximation ratio error by up to 72$\times$ compared to the unified stop condition-applied enhanced LMA-QAOA, and achieves equivalent approximation performance compared to the standard MA-QAOA.
}


\maketitle

\section{Introduction}

Quantum computing is expected to efficiently address certain problems that are hard to solve in classical computing \cite{acampora2024application, tennie2025quantum, majumder2024variational, vo2025q, rieffel2024assessing, incudini2024automatic, ovide2024scaling, kalis2023hybrid, chen2023quantum, kim2025qr}.  
Among various quantum-classical hybrid algorithms, the Quantum Approximate Optimization Algorithm (QAOA) has emerged as a prominent method for solving combinatorial optimization problems \cite{farhi2014quantum, liang2023hybrid, he2024parameter, hao2024end, he2023alignment}.
However, standard single-angle QAOA can require numerous circuit layers to achieve high-quality approximate solutions, increasing circuit depth and potentially limiting its practical deployment on near-term quantum devices.

\begin{figure}[h]
        \centering
        \includegraphics[width=\textwidth]{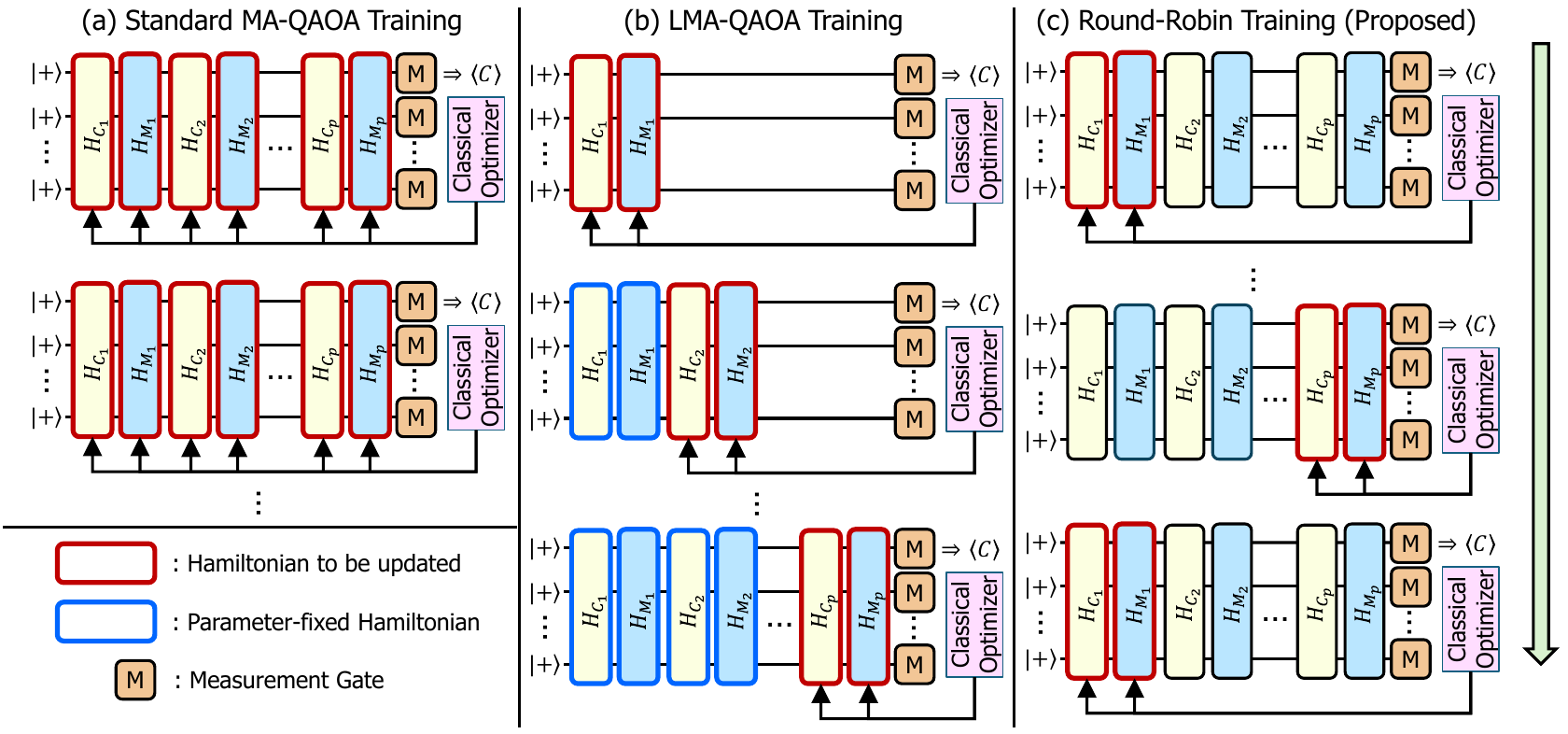}
    \hfill
    \caption{
        Comparison of MA-QAOA training strategies: 
        (a) Standard MA-QAOA trains all parameters simultaneously across all layers in each epoch. 
        (b) LMA-QAOA incrementally adds layers (and freezes previously optimized layers), reducing per-epoch optimization cost. 
        (c) The proposed Round-Robin training revisits each layer at a time in a cyclic fashion. 
    }
    \label{f11}
\end{figure}

To improve the expressibility \cite{holmes2022connecting} of QAOA circuits, Multi-Angle QAOA (MA-QAOA) was recently introduced, as shown in the Figure \ref{f11} (a) \cite{shi2022multiangle, srivastava2025improved, gaidai2024performance, herrman2022multi}.
Unlike standard QAOA, MA-QAOA assigns independent variational parameters to each Hamiltonian term, achieving superior approximation performance even with fewer layers.
Unfortunately, this increased expressibility \cite{holmes2022connecting} of MA-QAOA can increase the classical computational cost of parameter updates, as the number of parameters in each layer is required by the number of edges and nodes in the target graph.

To address this issue, Layerwise MA-QAOA (LMA-QAOA) has been proposed \cite{lavagna2024layerwise}.
In LMA-QAOA, the variational circuit is incrementally optimized one layer at a time, freezing previously trained parameters as new layers are added, as shown in Figure \ref{f11} (b).
This incremental approach reduces the number of parameters updated per epoch, lowering the computational overhead.
However, we observe that previously fixed parameters in early layers may hinder the ability to obtain precise solutions (on average of 3\% accuracy drop over the standard MA-QAOA).
This is because adding new layers changes the trainable parameter landscape when subsequent layers are introduced.

To understand the impact of previously fixed parameters, we first investigate whether parameters optimized in shallow circuits remain effective in deeper circuits, as shown in Section \ref{investigating}.
By analyzing the sufficiently trained Hamiltonian parameter configurations of QAOA circuits targeting equivalent graphs but with different numbers of layers, we could verify whether the LMA-QAOA's strategy of reusing the parameters is always efficient.
Our experiments in Section \ref{investigating} reveal that the configuration of mixer Hamiltonian parameters optimized when $p = 1$ (where $p$ denotes the number of layers) generally differ from those obtained at $p = 2$ or $p = 3$.
This characteristic indicates that early-trained parameters may not serve as optimal warm-start points for deeper layer scenarios.
Furthermore, LMA-QAOA gradually improves its approximation ratio as new layers are appended since the overall circuit expressibility increases, which may require more training steps compared to training deep layers together from the beginning.
Due to this, our evaluation using 6-node graphs demonstrates that the unified stop condition-applied LMA-QAOA requires 82.1\% more training steps than MA-QAOA, despite its lower per-step computational cost.

Nevertheless, we still note that partially updating QAOA parameters has a chance of reducing the classical computational overhead of MA-QAOA.
However, naively applying partial parameter updates may experience challenges from a qualitative perspective of training and inference of MA-QAOA circuits compared to collective updates of the entire layer. 
For efficient MA-QAOA training, this work addresses two questions:
(i) What is the optimal granularity for parameter updates per epoch, and 
(ii) How can we obtain accurate final cost function values while updating only a subset of parameters in each epoch?

Our experiments demonstrate that updating parameter groups less than layerwise granularity may be ineffective (as shown in the Section \ref{minimal}). 
The reason is that since the cost function calculation for gradient update is eventually calculated from the full edge and node information of the target graph, the minimal granularity should be a parameter set for the complete graph.
Meanwhile, we also observe that updating two layers per training step does not lead to a dramatically faster improvement in the approximation ratio compared to updating a single layer.
This implies that the parameter variations guided by the classical optimizer can be sufficiently reflected at the single-layer granularity level.
Utilizing appropriate layer retraining policies, layerwise updates are not slower in improving the cost function with training compared to full layer updates (i.e., standard MA-QAOA approach).
Thus, optimizing one complete layer per epoch can be an efficient granularity, balancing convergence efficiency and accuracy.

A potential concern for LMA-QAOA is that the previously fixed layers may hinder reaching a precise final cost function.
To address this, regularly retraining previously optimized layers enhances solution precision, since it allows for the exploitation of up-to-date parameters to adapt to changes introduced by additional layers.
Our evaluation confirms that retraining the layers sequentially in a sequential round-robin fashion or in random round-robin order can eventually reach a cost function equivalent to the final value provided by the standard MA-QAOA training (as shown in the Section \ref{orderlayer}).

To this end, we propose Orbit-QAOA (rOund RoBin layewIse mulTi-angle QAOA), a training framework designed for MA-QAOA training. 
The core idea of Orbit-QAOA is to revisit and retrain each layer cyclically in a round-robin fashion, rather than performing unidirectional grafting, as illustrated in Figure \ref{f11} (c). 
This cyclic optimization enables earlier layers to adapt dynamically to parameter changes introduced by subsequently added layers. 
Additionally, we observe that not all layers require equal optimization effort at each training epoch. 
To avoid redundant updates, Orbit-QAOA monitors the rate of change in the cost function at each parameter update step, and if this rate falls below a specified threshold, the layer is considered converged and excluded from subsequent optimization cycles. 
This selective freezing mechanism can identify stabilized layers, focusing computational resources specifically on layers requiring further refinement. 

Orbit-QAOA is evaluated comprehensively.
First, \Cref{f1} and \Cref{t1} show that increasing the number of layers in shallow circuits steadily enhances the Approximate Cut Ratio (ACR) while simultaneously accelerating convergence, even though Orbit updates only one layer per training step.
Next, \Cref{f2} and \Cref{t2} demonstrate that Orbit-QAOA achieves the identical performance of the ACR (approximated cut ratio) of 0.999 as standard MA-QAOA while reducing the number of training steps by 25.4\% and the overall runtime by 35\% on average.
Compared to the unified stop condition-applied LMA-QAOA, Orbit requires 2.45$\times$ fewer numbers of training steps to converge and achieves 3\% higher final ACR by avoiding premature layer freezing and retraining only the active layers.
As further shown in \Cref{t3}, Orbit's convergence advantage scales with the circuit depth: even for deeper QAOA circuits, Orbit continues to reduce the required number of training steps by an average of 25.5\% (and up to 59.3\%) while maintaining equivalent ACR to MA-QAOA.

In \Cref{f6} and \Cref{f7}, the analysis of the number of active layers reveals how Orbit freezes layers depending on the graph structure and training phase.
For simple cost Hamiltonians such as Power-Law graphs, the number of active layers rapidly decreases after ACR saturation, whereas for more complex graphs such as Barabási–Albert, layers remain active longer for further training, which shows Orbit-QAOA's adaptive control capability.
\Cref{f9} confirms that Orbit can maintain relatively higher ACR than MA-QAOA even when the activeness threshold $\varepsilon$ increases as Orbit's layerwise update provides more training opportunities in the final epoch.
Moreover, \Cref{f10} and \Cref{t5} show that Orbit could be generalized to the broader quantum alternating operator ansatz framework, maintaining identical ACR to MA-QAOA.
Finally, as shown in \Cref{f3} and \Cref{t4}, Orbit also accelerates single-angle (SA) QAOA convergence by approximately 20\% and achieves the same final ACR as the SA baseline, outperforming existing layerwise update approaches.

The key contributions of this paper are summarized as follows:

\begin{itemize}
    \item We find that fixed parameters of layers previously learned in LMA-QAOA may hinder obtaining precise final cost functions, as the addition of new layers in the QAOA circuit changes the training landscape.
    \item We propose Orbit-QAOA, a round-robin layerwise training strategy that iteratively revisits all layers of Multi-Angle QAOA (MA-QAOA), allowing previously optimized layers to adapt to newly added ones.
    \item We introduce a selective layer-freezing mechanism, which enables reducing unnecessary optimization steps by skipping layers that exhibit negligible cost function improvements, thereby enhancing training efficiency.
    \item Orbit-QAOA could offer a practical optimization strategy that can alleviate the classical computational overhead in training parameter-rich MA-QAOA circuits, contributing to more scalable hybrid quantum-classical workflows.
\end{itemize}

\section{Background and Motivation}

\subsection{Quantum Approximate Optimization Algorithm}

The quantum approximate optimization algorithm (QAOA) is a variational quantum algorithm designed to approximate solutions to combinatorial optimization problems \cite{farhi2014quantum}.  
One application using QAOA is the maximum cut (Max-Cut) problem, where the objective is to partition the set of nodes into two disjoint subsets such that the number of edges connecting the two subsets is maximized.  
Given a graph \( G = (V, E) \) with \( n = |V| \) nodes and \( m = |E| \) edges, the cost Hamiltonian is defined as

\[
C = \frac{1}{2} \sum_{(i, j) \in E} \left(1 - \sigma^i_z \sigma^j_z\right),
\]

where \( \sigma^k_z \) denotes the Pauli-Z operator applied to the \( k \)th qubit.  
This cost function encodes the expectation of the number of cuts as follows:

\[
C = \sum_{(i,j)\in E} \delta(z_i \neq z_j), \quad \text{with } z_i \in \{0,1\},
\]

where the diagonal Hamiltonian could be rewritten as follows:

\[
\delta(z_i \neq z_j) = \frac{1}{2}\left(1 - (-1)^{z_i \oplus z_j}\right) = \frac{1}{2}\left(1 - \langle \sigma^i_z \sigma^j_z \rangle \right).
\]

The observable of the classical objective function can be represented by the Ising Hamiltonian structure above.
The solution to the Max-Cut problem can be encoded in a computational basis state \( \lvert \psi \rangle \), where each bit \( s_i \in \{0, 1\} \) represents the subset assignment of node \( i \).  
QAOA begins by initializing the qubit register in a uniform superposition state:

\[
\lvert \psi_0 \rangle = H^{\otimes n} \lvert 0 \rangle^{\otimes n} = \lvert + \rangle^{\otimes n}.
\]

Then, alternating unitaries of the cost Hamiltonian and a mixer Hamiltonian are applied in sequence.  
For a circuit with \( p \) layers, the variational ansatz is expressed as

\[
\lvert \psi_p(\boldsymbol{\gamma}, \boldsymbol{\beta}) \rangle = \left(\prod_{l=1}^{p} U_M(\beta_l) U_C(\gamma_l) \right) \lvert \psi_0 \rangle,
\]

where the cost and mixer unitaries are defined by

\[
U_C(\gamma) = \exp\left(-i \gamma C\right), \quad U_M(\beta) = \exp\left(-i \beta \sum_{j=1}^{n} \sigma^j_x \right).
\]

Note that the cost unitary can be expanded as a product of two-qubit ZZ interaction operators:

\[
U_C(\gamma) = \prod_{(i,j) \in E} \exp\left( i \frac{\gamma}{2} \sigma^i_z \sigma^j_z \right) \cdot \exp\left( -i \frac{\gamma}{2} \right),
\]

up to a global phase that does not affect measurement outcomes.

Here, \( \boldsymbol{\gamma} = (\gamma_1, \ldots, \gamma_p) \) and \( \boldsymbol{\beta} = (\beta_1, \ldots, \beta_p) \) are the variational parameters to be optimized.  
In this setting, the parameters within the cost and mixer Hamiltonians at each layer are shared across all terms.  
The expectation value of the cost Hamiltonian is measured as

\[
\langle C \rangle = \langle \psi_p(\boldsymbol{\gamma}, \boldsymbol{\beta}) \lvert C \rvert \psi_p(\boldsymbol{\gamma}, \boldsymbol{\beta}) \rangle,
\]

and a classical optimizer updates the parameters to maximize this value.  
The hybrid process of alternating quantum circuit execution and classical parameter update continues until convergence of the cost function is achieved.

\subsection{Multi-Angle QAOA}

While the standard (single-angle) QAOA provides a compact parameterization of the cost and mixer unitaries, its limited expressibility \cite{holmes2022connecting} may necessitate deeper circuits to capture complex solution landscapes.  
To address this, the Multi-Angle QAOA (MA-QAOA) extends the standard QAOA by assigning distinct variational parameters to each term in the Hamiltonian \cite{herrman2022multi}.  
This fine-grained parameterization allows the representational increase of the variational ansatz while preserving the alternating structure of cost and mixer operators.
In MA-QAOA, the cost and mixer unitaries for each layer \( l \in \{1, \dots, p\} \) can be written as

\[
U_{C}^{(l)} = \prod_{(i,j) \in E} \exp\left(i \gamma_{l}^{(i,j)} \sigma^i_z \sigma^j_z \right), \quad
U_{M}^{(l)} = \prod_{j=1}^{n} \exp\left(-i \beta_{l}^{(j)} \sigma^j_x \right),
\]

where \( \gamma_{l}^{(i,j)} \in \mathbb{R} \) and \( \beta_{l}^{(j)} \in \mathbb{R} \) are the variational parameters assigned to edge \( (i,j) \) and qubit \( j \) in layer \( l \), respectively.  
The resulting MA-QAOA state is constructed as

\[
\lvert \psi_p(\boldsymbol{\Gamma}, \boldsymbol{B}) \rangle = 
\left( \prod_{l=1}^{p} U_{M}^{(l)} U_{C}^{(l)} \right) \lvert \psi_0 \rangle,
\]

where the parameter sets are defined as
\[
\boldsymbol{\Gamma} = \left\{ \gamma_{l}^{(i,j)} \;\middle|\; 1 \leq l \leq p,\; (i,j) \in E \right\}, \quad
\boldsymbol{B} = \left\{ \beta_{l}^{(j)} \;\middle|\; 1 \leq l \leq p,\; 1 \leq j \leq n \right\},
\]
with \( \boldsymbol{\Gamma} \in \mathbb{R}^{pm} \) and \( \boldsymbol{B} \in \mathbb{R}^{pn} \).  
Here, \( E \) denotes the set of edges in the target graph with \( m = |E| \), and \( n \) is the number of nodes.

The standard (single-angle) QAOA is recovered as a special case of MA-QAOA when the parameters are shared across all corresponding terms:

\[
\text{Standard QAOA} \iff
\gamma_{l}^{(i,j)} = \gamma_l, \quad \beta_{l}^{(j)} = \beta_l \quad \forall (i,j),\; j.
\]

This inclusion relationship implies that MA-QAOA generalizes QAOA in terms of expressibility, enabling the variational circuit to explore a broader subspace of the Hilbert space.  
The increased flexibility, however, comes with a higher classical optimization cost, as the total number of parameters scales as \( O(pm + pn) \) for a problem with \( m \) edges and \( n \) qubits.  
This may result in longer training time.  
This leads to a greater computational overhead on the classical side compared to Single-Angle QAOA, which requires only one cost and one mixer Hamiltonian's parameters per layer (i.e., $O(2p)$).
Nonetheless, it has been shown that the approximation ratio of MA-QAOA converges to the optimal value as \( p \to \infty \), and MA-QAOA achieves equal or better performance than QAOA at the same depth~\cite{herrman2022multi}.  
For instance, in the MaxCut problem on star graphs, a single layer of MA-QAOA achieves an approximation ratio of 1, whereas standard QAOA approaches 0.75 asymptotically \cite{herrman2022multi}.

\subsection{Layerwise-Multi-Angle QAOA}

The Layerwise-Multi-Angle QAOA (LMA-QAOA) aims to reduce the optimization burden of MA-QAOA by applying a parameter-fixing strategy in a sequential, layerwise manner \cite{lavagna2024layerwise}. 
While MA-QAOA maximizes expressibility by assigning independent parameters to each edge and qubit in every layer, this may come at the cost of optimization tractability due to the explosion in the number of parameters. 
LMA-QAOA addresses this by progressively growing the circuit depth and only optimizing the new parameters introduced in each layer while freezing the previously optimized ones.
Let \( H(\sigma) \) denote the cost Hamiltonian associated with a combinatorial optimization problem in the Ising form:

\[
H(\sigma) = \sum_{i \in V} \theta_i \sigma^i_z + \sum_{(i,j) \in E} \theta_{ij} \sigma^i_z \sigma^j_z,
\]

where \( \sigma^i_z \) denotes the Pauli-Z operator acting on qubit \( i \), and \( \theta_i \), \( \theta_{ij} \) are the local and pairwise interaction weights, respectively. 
The initial quantum state is prepared as

\[
\lvert \psi_0 \rangle = \frac{1}{\sqrt{2^n}} \sum_{z \in \{0,1\}^n} \lvert z \rangle.
\]

For a given layer \( l \), the cost and mixer unitaries in MA-QAOA form are defined as:

\[
U_C(\gamma_l) = \exp\left(-i \sum_{(i,j) \in E} \gamma^{(l)}_{i,j} \sigma^i_z \sigma^j_z\right), \quad
U_M(\beta_l) = \exp\left(-i \sum_{j=1}^{n} \beta^{(l)}_j \sigma^j_x\right),
\]

where \( \gamma^{(l)}_{i,j} \in [0, 2\pi) \) and \( \beta^{(l)}_j \in [0, \pi) \) are layer-specific parameters. 
In LMA-QAOA, the total variational state at layer \( l \) is recursively constructed as:

\[
\lvert \psi_l(\gamma_1, \ldots, \gamma_l, \beta_1, \ldots, \beta_l) \rangle =
U_M(\beta_l) U_C(\gamma_l) \cdots U_M(\beta_1) U_C(\gamma_1) \lvert \psi_0 \rangle.
\]

However, unlike MA-QAOA, the classical optimization at layer \( l \) only updates \( \gamma_l \) and \( \beta_l \), while all previous parameters \( \gamma_1, \ldots, \gamma_{l-1} \) and \( \beta_1, \ldots, \beta_{l-1} \) are fixed to their optimized values \( \gamma^*_1, \ldots, \gamma^*_{l-1} \) and \( \beta^*_1, \ldots, \beta^*_{l-1} \), respectively. That is, for each step \( l > 1 \), the optimization is carried out with:

\[
\gamma^{(1:l)} = (\gamma^*_1, \ldots, \gamma^*_{l-1}, \gamma_l), \quad
\beta^{(1:l)} = (\beta^*_1, \ldots, \beta^*_{l-1}, \beta_l).
\]

The energy expectation is then evaluated as follows:

\[
F_l(\gamma_l, \beta_l) = \langle \psi_l(\cdot) \rvert C (\sigma) \lvert \psi_l(\cdot) \rangle,
\]

and used to optimize only the parameters at layer \( l \). This sequential approach reduces the effective dimensionality of the parameter space at each optimization step from \( O(pm + pn) \) in MA-QAOA to \( O(m + n) \) in LMA-QAOA, where \( m = |E| \) and \( n = |V| \).

\subsection{Trained Parameter Analysis across Layers in MA-QAOA} \label{investigating}

LMA-QAOA efficiently reduces the per-epoch optimization overhead by training only one layer at a time~\cite{lavagna2024layerwise}.  
According to our evaluations with a 6-node target graph of Erdős–Rényi model \cite{erdHos2013spectral}, LMA-QAOA achieves a \(2.27\times\) reduction in training time per step compared to standard MA-QAOA.  
However, despite this advantage, the total number of training steps increases by \(5.21\times\), resulting in an overall training simulation cost that is \(2.29\times\) higher than that of MA-QAOA.

This increase in the total number of training steps can be attributed to the fact that when a new layer is added in MA-QAOA, the underlying parameter landscape itself changes.  
Previously trained layers optimized under shallower circuit configurations may no longer serve as effective warm-starts in the modified optimization space.  
Since MA-QAOA assigns distinct parameters to each operator, its parameter space is extremely high-dimensional, making it impractical to directly visualize how the training landscape evolves as the circuit depth increases.
To investigate the change of the parameter landscape indirectly, we compare the trained parameters across MA-QAOA circuits with the same target graph but with different layers.  
Table \ref{t0} shows the sufficiently trained mixer Hamiltonian parameters across all layers of MA-QAOA circuits for the 6-node Sherrington-Kirkpatrick (SK) model as the number of layers \( p \) increases.

\begin{table}[h]
\caption{
Comparison of the mixer Hamiltonian parameters across all layers for the trained 6-node SK-model multi-angle QAOA circuits (p = 1, 2, and 3) 
}\label{t0}
\centering
\begin{tabular}{@{\extracolsep{\fill}}cc|cccccc@{}}
\toprule
$p$ & Layer & $\beta_0$ & $\beta_1$ & $\beta_2$ & $\beta_3$ & $\beta_4$ & $\beta_5$ \\
\midrule
\multirow{1}{*}{$p=1$} & Layer 0 & -0.09 & 0.11 & 0.49 & -0.11 & -0.53 & -0.61 \\
\midrule
\multirow{2}{*}{$p=2$} 
& Layer 0 & 0.49 & 0.55 & 0.43 & 0.02 & 0.02 & -0.03 \\
& Layer 1 & 0.25 & 0.24 & 0.29 & 0.27 & 0.01 & 0.04 \\
\midrule
\multirow{3}{*}{$p=3$} 
& Layer 0 & 0.03 & 0.43 & -0.05 & 0.36 & -0.22 & -0.15 \\
& Layer 1 & -0.29 & -0.11 & -0.19 & 0.23 & -0.16 & -0.08 \\
& Layer 2 & -0.19 & -0.17 & -0.12 & -0.22 & -0.15 & -0.14 \\
\botrule
\end{tabular}
\end{table}

As observed in Table \ref{t0}, the trained mixer Hamiltonian parameters' configurations of the 0-th layer when \( p = 1 \) do not resemble those of all layers when \( p = 2 \) and \( p = 3 \).  
These results suggest that the configuration of the sufficiently trained parameters in one layer depends on the number of layers.
A layer trained early in a shallow circuit may no longer represent an optimal setting once subsequent layers are appended.
Furthermore, note that increasing the number of layers inherently enhances the expressibility of the QAOA circuit.
Accordingly, in LMA-QAOA, additional training iterations may be required to reflect this increased expressibility as new layers are grafted.
This implies that the overall convergence in LMA-QAOA can be slower than in MA-QAOA, which trains all layers jointly from the beginning.
Due to these, although the higher expressibility of MA-QAOA enables LMA-QAOA to asymptotically approach the theoretical approximation ratio when a sufficient number of layers are provided, the number of steps required for convergence by LMA-QAOA would rather be higher than that by MA-QAOA.  

Therefore, reusing trained parameters from shallower circuit configurations, as done in LMA-QAOA, may not always accelerate training the parametrized circuits.  
These observations motivate the design of Orbit-QAOA, which preserves the cost-efficiency of layerwise training by updating only one layer per epoch but allows all active layers to be re-optimized in a round-robin fashion.  
By enabling earlier layers to adapt to the changes induced by newly added layers, Orbit-QAOA better exploits the evolving parameter landscape, achieving faster convergence without sacrificing the expressibility of multi-angle parameterization.

\section{Orbit-QAOA Design} \label{orbit_design}

This section explains the proposed Orbit-QAOA approach.
The proposed Orbit-QAOA trains each layer in a round-robin manner and tracks the change in the cost function (\( \Delta C \)) at each epoch to identify stabilized layers and exclude them from subsequent training iterations.  

\subsection{Round-robin Layerwise Training}

Orbit-QAOA employs a round-robin strategy that cyclically revisits all layers until their convergence conditions are individually met.  
In each training epoch, only one layer is updated, which preserves the per-step optimization cost the same as that of LMA-QAOA.  
For each step \( t \), Orbit-QAOA selects a layer \( l \in \mathcal{A}^{(t)} \), and optimizes only its corresponding parameters \( \gamma_l \) and \( \beta_l \), while keeping all others fixed. 
The objective for this local update is defined as

\[
(\gamma_l^{(t+1)}, \beta_l^{(t+1)}) = \arg\min_{\gamma_l, \beta_l} C(\gamma^{(t)}_{<l}, \gamma_l, \gamma^{(t)}_{>l};\; \beta^{(t)}_{<l}, \beta_l, \beta^{(t)}_{>l}),
\]

where \( \gamma^{(t)}_{<l} = (\gamma_1^{(t+1)}, \ldots, \gamma_{l-1}^{(t+1)}) \), and \( \gamma^{(t)}_{>l} = (\gamma_{l+1}^{(t)}, \ldots, \gamma_p^{(t)}) \) denote the most recent values available for other layers.  
The current step reuses parameters updated in previous rounds (i.e., \( \gamma_k^{(t+1)} \) for \( k < l \)), making the update of layer \( l \) adaptive to the evolution of the circuit state across the previous epoch.

The quantum state at each step is expressed as

\[
\lvert \psi_p(\gamma, \beta) \rangle = \left( \prod_{k=p}^1 U_M(\beta_k) U_C(\gamma_k) \right) \lvert + \rangle^{\otimes n},
\]

and the cost function is given by the expectation

\[
C(\gamma, \beta) = \langle \psi_p(\gamma, \beta) \rvert C \lvert \psi_p(\gamma, \beta) \rangle.
\]

Although only one layer is updated per training step, the round-robin schedule allows every layer to be revisited after others have changed.  
Let layer \( l \) be layerwisely updated at step \( t \), using the fixed parameters \( \gamma_k^{(t)} \), \( \beta_k^{(t)} \) from all other layers \( k \neq l \).:

\[
\nabla_{\gamma_l}^{(t)} C = \frac{\partial}{\partial \gamma_l} \langle \psi_p(\gamma, \beta) \rvert C \lvert \psi_p(\gamma, \beta) \rangle, \quad
\nabla_{\beta_l}^{(t)} C = \frac{\partial}{\partial \beta_l} \langle \psi_p(\gamma, \beta) \rvert C \lvert \psi_p(\gamma, \beta) \rangle.
\]

Thus, each layer is given the opportunity to adapt to changes made by other layers during previous updates.  
This allows each layer in Orbit-QAOA to adapt its parameters in response to variations introduced by other layers in previous steps.  
This inter-round dependency could allow for trainable layers to correct earlier choices of parameter updates that may not have become an optimal selection from the perspective of the full circuit.  
It also enables the algorithm to further exploit the expressibility of the MA-QAOA ansatz than the LMA-QAOA since each layer can iteratively align its contribution with the evolving structure of the entire quantum state.

\subsection{Skipping Learning of Stabilized Layers} \label{skip}

In the round-robin layerwise training approach proposed in this work, it is not necessary to include all layers in the training schedule every time.
In other words, some sufficiently stabilized layers can improve training efficiency by excluding them from training scheduling.
Orbit-QAOA introduces a mechanism for dynamically excluding layers that have converged.  
At each round, the algorithm evaluates whether the currently updated layer exhibits meaningful improvement in the cost function and removes it from future updates if not.  
This approach reduces redundant computation and focuses optimization resources toward layers that remain under-trained.

Let \( C^{(t)} \) denote the value of the cost function at training step \( t \), evaluated as the expectation of the problem Hamiltonian:

\[
C^{(t)} = \langle \psi_p(\gamma^{(t)}, \beta^{(t)}) \rvert C \lvert \psi_p(\gamma^{(t)}, \beta^{(t)}) \rangle.
\]

For a given layer \( l \in \mathcal{A}^{(t)} \), we define its contribution to improvement in the expected number of cuts as

\[
\Delta C_l^{(t)} = C_l^{\text{after}} - C_l^{\text{before}},
\]

where \( C_l^{\text{before}} \) and \( C_l^{\text{after}} \) represent the cost function values immediately before and after updating layer \( l \) at step \( t \), respectively.  
If the improvement magnitude satisfies

\[
|\Delta C_l^{(t)}| < \varepsilon,
\]

for a small positive threshold \( \varepsilon \), then the layer is considered stabilized and removed from the active set:

\[
\mathcal{A}^{(t+1)} = \mathcal{A}^{(t)} \setminus \{l\}.
\]

This layerwise criterion allows the training process to continue for non-converged parts of the circuit while halting updates to layers that no longer provide benefit.  
As training proceeds, the number of active layers \( |\mathcal{A}^{(t)}| \) decreases, and the round-robin scheduler operates over a shrinking set.

\subsection{Activeness Threshold Setting} \label{convergence}

A critical component of the layer-skipping mechanism in Orbit-QAOA is the appropriate selection of the activeness threshold \( \varepsilon \).  
The threshold must be set sufficiently small to ensure that a layer is excluded from the round-robin schedule only after meaningful optimization progress has been exhausted.  
The setting of \( \varepsilon \) should consider the minimal quantized change in the cost function induced by the sampling resolution of QAOA circuit evaluations.

Suppose the QAOA circuit is evaluated with \( S \) sampling shots \cite{kim2024distribution} at the current epoch.  
Among these samples, let \( m \) shots for several bitstrings correspond to a cut count of \( i \) for the target graph, and \( n \) shots for several bitstrings correspond to a cut count of \( i+1 \).  
The expectation of the number of cuts can be expressed as

\[
\hat{C}^{(t)} = \frac{1}{S} \left( m \times i + n \times (i+1) + \sum_{k \notin \{i, i+1\}} c_k^{(t)} \times k \right),
\]

where \( c_k^{(t)} \) denotes the number of shots resulting in a cut count of \( k \) at epoch \( t \).

Now assume that after parameter updates at the next epoch, the QAOA circuit is again sampled with \( S \) shots, resulting in \( m-1 \) shots corresponding to a cut count of \( i \) and \( n+1 \) shots corresponding to a cut count of \( i+1 \), while the distribution for other cut counts remains unchanged.  
The updated expectation of the number of cuts becomes

\[
\hat{C}^{(t+1)} = \frac{1}{S} \left( (m-1) \times i + (n+1) \times (i+1) + \sum_{k \notin \{i, i+1\}} c_k^{(t)} \times k \right).
\]

The change in the expected cut count between two consecutive epochs is therefore given by

\[
\Delta C = \hat{C}^{(t+1)} - \hat{C}^{(t)} = \frac{1}{S} \left( (-1) \times i + (+1) \times (i+1) \right) = \frac{1}{S}.
\]

Thus, under the given sampling condition, the minimal quantized change in the cost function is

\[
\Delta C = \frac{1}{S}.
\]

Setting \( \varepsilon \) to be slightly larger than \( 1/S \) ensures that a layer is excluded from the round-robin schedule only when parameter updates fail to induce meaningful changes beyond the quantization limit determined by the sampling resolution.
In this work, the threshold $ \varepsilon $ is set once before training and remains fixed throughout the entire optimization process.

In particular, when employing adaptive optimizers such as the Adaptive Gradient Algorithm (Adagrad \cite{duchi2011adaptive}), where the effective learning rate naturally decreases over time due to the accumulation of past gradient information, it is critical to adopt a conservative threshold setting.  
Otherwise, layers might be skipped not due to the convergence but simply because the classical optimizer's step size has diminished.

Based on our empirical observations, when \( S = 1,024 \) shots are used for evaluating each QAOA circuit, setting \( \varepsilon = 0.001 \) (which is slightly bigger than 1/$S$) could provide a reasonable balance between convergence sensitivity and optimization stability.  
This configuration allows Orbit-QAOA to determine layer-freezing when further training of a layer is no longer productive while avoiding premature exclusion caused by the statistical limitations of finite-shot evaluations.

\subsection{Algorithm Implementation} \label{overview}

Let the QAOA circuit be composed of \( p \) layers. 
The variational parameters are defined as \( \gamma = (\gamma_1, \ldots, \gamma_p) \) and \( \beta = (\beta_1, \ldots, \beta_p) \), where each pair \( (\gamma_l, \beta_l) \) means the cost and mixer unitaries of layer \( l \), respectively. 
At each training step \( t \), Orbit-QAOA maintains a subset \( \mathcal{A}^{(t)} \subseteq \{1, \ldots, p\} \), referred to as the \textit{active layer set}, which contains the indices of layers that are still subject to optimization.
Orbit-QAOA evaluates whether each layer should remain active by measuring the change in the cost function value, defined as the expected number of cuts represented by the Hamiltonian expectation

\[
C(\gamma, \beta) = \langle \psi_p(\gamma, \beta) \rvert C \lvert \psi_p(\gamma, \beta) \rangle.
\]

For a given layer \( l \in \mathcal{A}^{(t)} \), only its associated parameters \( \gamma_l \) and \( \beta_l \) are optimized, while the remaining parameters are held fixed. 
The improvement in the cost function is computed as

\[
\Delta C_l^{(t)} = C_l^{\text{after}} - C_l^{\text{before}},
\]

where \( C_l^{\text{before}} \) and \( C_l^{\text{after}} \) denote the expected number of cuts immediately before and after updating layer \( l \) at step \( t \). 
If the magnitude of this change satisfies \( |\Delta C_l^{(t)}| < \varepsilon \), the layer is considered to have stabilized and is removed from the active set:

\[
\mathcal{A}^{(t+1)} = \mathcal{A}^{(t)} \setminus \{l\}.
\]

\subsubsection{Orbit-QAOA Training Implementation} \label{orbit-implement}

\begin{algorithm}[H]
\caption{Orbit-QAOA Training Process}
\label{a1}
\begin{algorithmic}[1]
\State Initialize $\boldsymbol{\gamma} \in \mathbb{R}^{p \times m}$ and $\boldsymbol{\beta} \in \mathbb{R}^{p \times n}$ with small random values
\State Set \texttt{requires\_grad = True} for all entries
\State Initialize classical optimizer \texttt{opt}
\State $\mathcal{A} \gets \{0, 1, \dots, p-1\}$ \Comment{Set of active layers}
\State $\texttt{hist} \gets [\ ]$
\While{$\mathcal{A} \neq \emptyset$}
  \ForAll{$\ell \in$ \Call{Sorted}{$\mathcal{A}$}}
    \For{$p = 0$ to $p-1$}
      \State Set $\texttt{gradient} = (\texttt{True} \text{ if } p = \ell \text{ else } \texttt{False})$ for $\boldsymbol{\gamma}[p]$ and $\boldsymbol{\beta}[p]$
    \EndFor
    \State $C_{\text{before}} \gets$ \Call{Objective}{$\boldsymbol{\gamma}, \boldsymbol{\beta}$}
    \State $(\boldsymbol{\gamma}, \boldsymbol{\beta}) \gets$ \Call{opt.step}{Objective, $\boldsymbol{\gamma}, \boldsymbol{\beta}$}
    \State $C_{\text{after}} \gets$ \Call{Objective}{$\boldsymbol{\gamma}, \boldsymbol{\beta}$}
    \State Append $C_{\text{after}}$ to \texttt{hist}
    \State $\Delta \gets C_{\text{after}} - C_{\text{before}}$
    \If{$|C_{\text{after}} - C_{\text{before}}| < \varepsilon$}
      \State $\mathcal{A} \gets \mathcal{A} \setminus \{\ell\}$
    \EndIf
  \EndFor
\EndWhile
\State \Return \texttt{hist}
\end{algorithmic}
\end{algorithm}

Algorithm \ref{a1} describes the implementation of Orbit-QAOA, which performs round-robin training of multi-angle QAOA circuits with layer skipping.  
The algorithm begins by initializing the trainable parameters \( \boldsymbol{\gamma} \in \mathbb{R}^{p \times m} \) and \( \boldsymbol{\beta} \in \mathbb{R}^{p \times n} \), where \( p \) is the number of QAOA layers, \( m = |E| \) is the number of edges in the target graph, and \( n = |V| \) is the number of qubits.  
Each parameter is initialized with a small random value, and gradient tracking is enabled.
These randomized initials allow the training of all Hamiltonian operators to begin at points with a slight gradient near the starting point in the training landscape rather than starting at all zero angles, thereby facilitating the initial ACR improvements of QAOA training \cite{bergholm2018pennylane}.
A classical optimizer is used to perform parameter updates, and the set of active layers \( \mathcal{A} \subseteq \{0, \dots, p-1\} \) is maintained throughout training.

Training proceeds iteratively while \( \mathcal{A} \) is non-empty.  
At each step, the algorithm traverses the active layers \( \ell \in \mathcal{A} \) in sorted order.  
For each selected layer \( \ell \), only its corresponding entries in \( \boldsymbol{\gamma} \) and \( \boldsymbol{\beta} \) are set to require gradients (i.e., activate the layer), while all others are temporarily frozen.  
The cost function is evaluated immediately before and after performing a single optimizer update.  
Let \( C_{\text{before}} \) and \( C_{\text{after}} \) denote the objective values before and after updating layer \( \ell \), respectively, and define the progress as \( \Delta = C_{\text{after}} - C_{\text{before}} \).  
If \( |\Delta| < \varepsilon \), the layer is deemed converged and removed from the active set \( \mathcal{A} \).

The training step count is incremented after every layer update, and the value of \( C_{\text{after}} \) is appended to the training history \texttt{hist}.  
Once all layers converge, the optimization process terminates, and the history of cost values is returned.

\section{Observation} \label{observe}

\subsection{Minimal Quantum to be Updated for Efficient Training} \label{minimal}

\begin{figure}[h]
    \centering
    \begin{subfigure}[t]{0.49\textwidth}
        \centering
        \includegraphics[width=\textwidth]{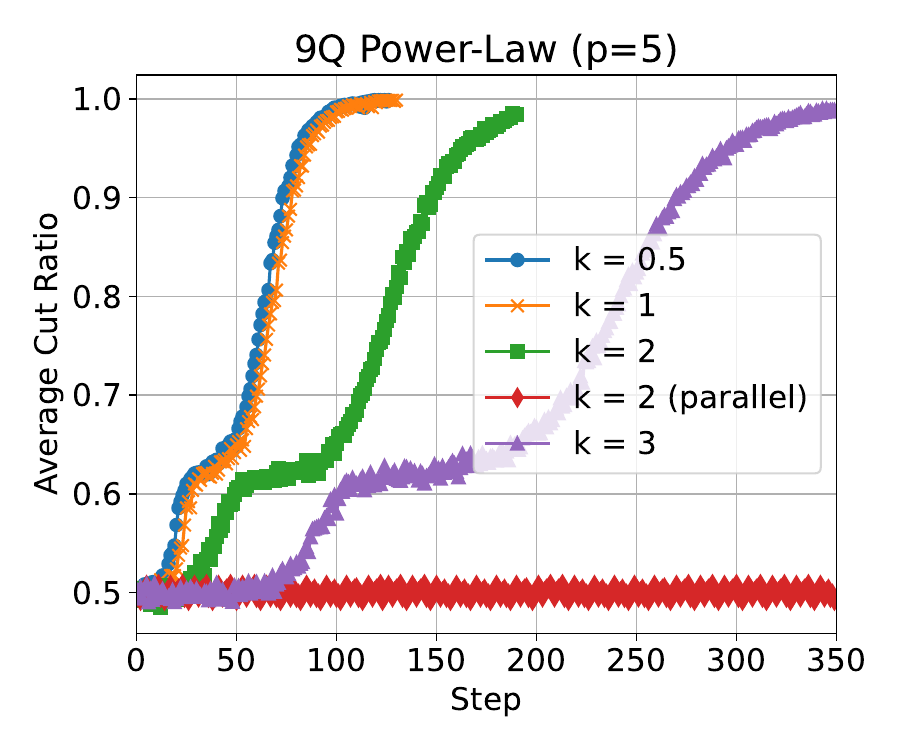}
    \end{subfigure}
    \hfill
    \begin{subfigure}[t]{0.49\textwidth}
        \centering
        \includegraphics[width=\textwidth]{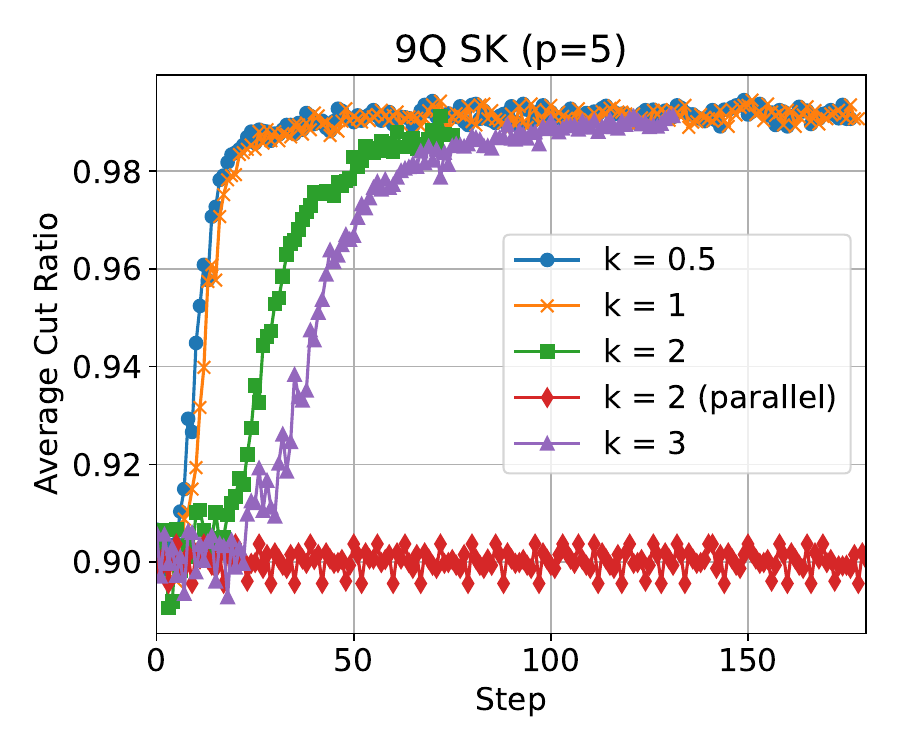}
    \end{subfigure}
    \caption{
    Training performance of  9-qubit 5-layer MA-QAOA circuits on graphs for Power-Law and Sherrington–Kirkpatrick models under sublayer-partitioned optimization by round-robin training approach.  
    The value \( k \) denotes the number of sublayers into which each QAOA layer is partitioned.  
    In other words, 1/$k$ of the single-layer's parameters are updated per step.
    }
    \label{f4}
\end{figure}

In this section, we investigate whether partitioning the QAOA circuit into finer or coarser granularities than the standard layerwise structure can provide additional benefits for training efficiency.
We consider a sublayer-partitioned variant of MA-QAOA, where each layer’s cost and mixer Hamiltonian parameters are divided into $k$ disjoint sublayers.
At each training step, only one sublayer is updated according to a round-robin scheduling policy.
When $k = 1$, each sublayer corresponds to a full QAOA layer, which is equivalent to the Orbit-QAOA approach proposed in this work.
When $k = 0.5$, two out of five layers are sequentially selected and updated per training step.
This configuration performs updates at a finer granularity than MA-QAOA but is coarser than the layerwise update strategy used in Orbit-QAOA.
When $k = 2$, half of the parameters within a single layer are updated at each step, whereas $k = 3$ corresponds to updating one-third of a layer per step.
In the case of $k = 2$ (parallel), half of a layer’s parameters are divided into two groups and updated in parallel during each training step, allowing partial layer updates to proceed concurrently.

\textbf{$k = 1$ VS. $k = 0.5$}:
As shown in \Cref{f4}, for both Power-Law and Sherrington–Kirkpatrick target graphs, the ACR growth curves are nearly identical when updating one layer per step ($k=1$) and when updating two layers per step ($k=0.5$).
Although the $k=1$ scenario updates only half as many parameters per step, its ACR improvement remains almost the same, which may appear counterintuitive.
However, this observation indicates that the contribution of parameter updates from a single layer is sufficient to reflect meaningful ACR improvement provided by the classical optimizer.
This finding implies that updating multiple layers simultaneously may not yield additional benefits dramatically.
The fact that updating one layer can achieve a comparable increase in ACR to that of full-circuit updates could also be confirmed by comparing training results of Orbit-QAOA and MA-QAOA shown in \Cref{f2}.
When similar ACR improvements could be achieved, the configuration that imposes less computational load on the classical optimizer would be preferable.
Therefore, from the viewpoint of QAOA circuit training efficiency, $k=1$ is more advantageous than $k=0.5$.

\textbf{$k = 1$ vs. $k = 2$ or $3$}:
As shown in \Cref{f4}, both Power-Law and Sherrington–Kirkpatrick target graphs exhibit slower convergence as the number of sublayers $k$ increases.
When the parameters are trained with finer granularity (fewer parameters in one training batch), it is natural to observe slower convergence because the parameters are no longer updated globally.
This can also be explained by how the cost function is evaluated.
While the cost function is related to all the parameters associated with a single complete layer, only a subset of those parameters is updated at each step when $k > 1$.
That is, the gradient is computed with respect to the entire layer, but only the selected sublayer receives parameter updates.
As $k$ increases, this imbalance becomes more pronounced, leading to slower learning due to underutilization of the available gradient information.
Modifying the cost function to depend only on the subset of the circuit corresponding to the selected sublayer would redefine the problem on a partial graph, however, which is not semantically valid since it alters the original optimization objective.

\textbf{$k = 2$ vs. $k = 2$ (parallel)}:
In the Power-Law graph shown in \Cref{f4}, the $k = 1$ configuration (orange line) achieves an ACR close to 1 at around 130 steps, whereas the $k = 2$ configuration (green line) reaches a similar ACR at approximately 190 steps.
This degradation in convergence speed is less than twofold.
Thus, one might naturally wonder whether updating half of a layer’s parameters in parallel could provide a speed advantage over $k = 1$.
The $k = 2$ (parallel) scenario explores this idea of dividing each layer’s parameters into two halves, updating them concurrently, and then merging the results.
Unfortunately, unlike the sequential $k = 2$ case, parallel parameter updates could not establish a consensus of agreement on the gradient improvement direction between the two concurrent executions.
This may lead to no continuous progress in ACR improvements as shown in \Cref{f4}.
In contrast, although the sequential $k = 2$ configuration converges more slowly, its sublayers maintain dependency between parameter updates, allowing the classical optimizer to continue guiding training toward higher ACR values.
In single-angle QAOA (SA-QAOA), parallel updates across different layers may indeed be efficient under the parameter-shift rule \cite{mitarai2018quantum}, since each parameterized operator $e^{-i\theta P}$ contributes independently to gradient evaluation, allowing $\langle C \rangle$ to be computed in parallel via twice evaluations of $\theta \pm \pi/2$.
However, performing parallel updates within the same layer in MA-QAOA is inherently difficult because their parameters are mutually interdependent.
Each simultaneously partial update within a layer must assume that the remaining parameters of that layer remain fixed as before.
As a result, it becomes challenging to achieve coherent improvements among the interdependent sub-parameters within the same layer, where the Hamiltonian is collectively defined.

Therefore, we can conclude that the single QAOA layer is the minimal quantum of parameter granularity that provides efficient training performance.

\subsection{Selecting the Order of Layers to Be Updated} \label{orderlayer}

\begin{figure}[h]
    \centering
    \begin{subfigure}[t]{0.32\textwidth}
        \centering
        \includegraphics[width=\textwidth]{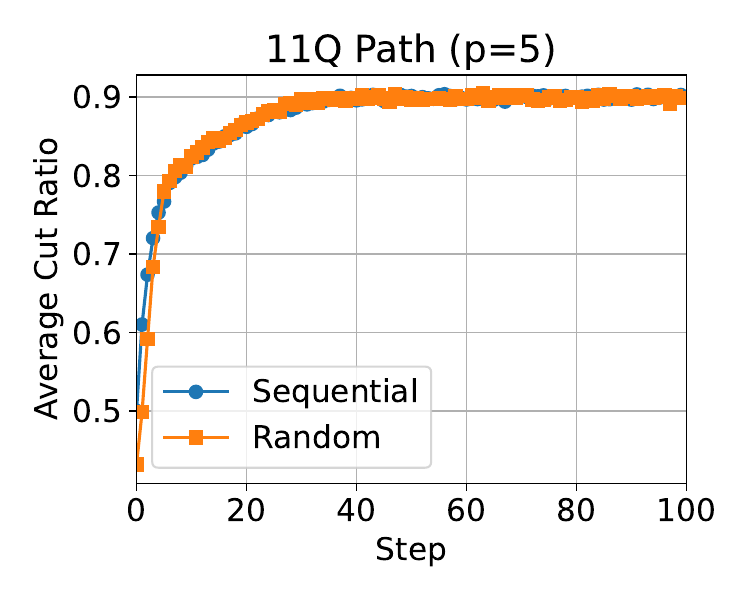}
    \end{subfigure}
    \hfill
    \begin{subfigure}[t]{0.32\textwidth}
        \centering
        \includegraphics[width=\textwidth]{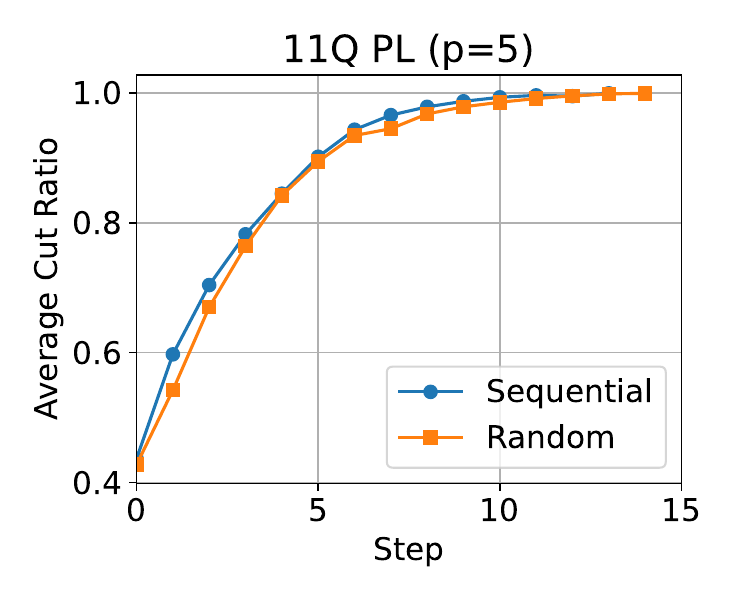}
    \end{subfigure}
    \hfill
    \begin{subfigure}[t]{0.32\textwidth}
        \centering
        \includegraphics[width=\textwidth]{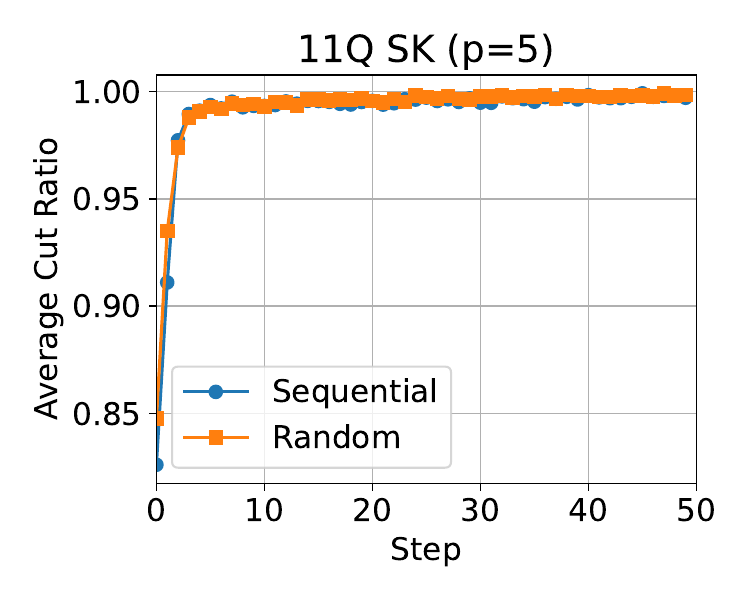}
    \end{subfigure}
    \hfill
    \begin{subfigure}[t]{0.32\textwidth}
        \centering
        \includegraphics[width=\textwidth]{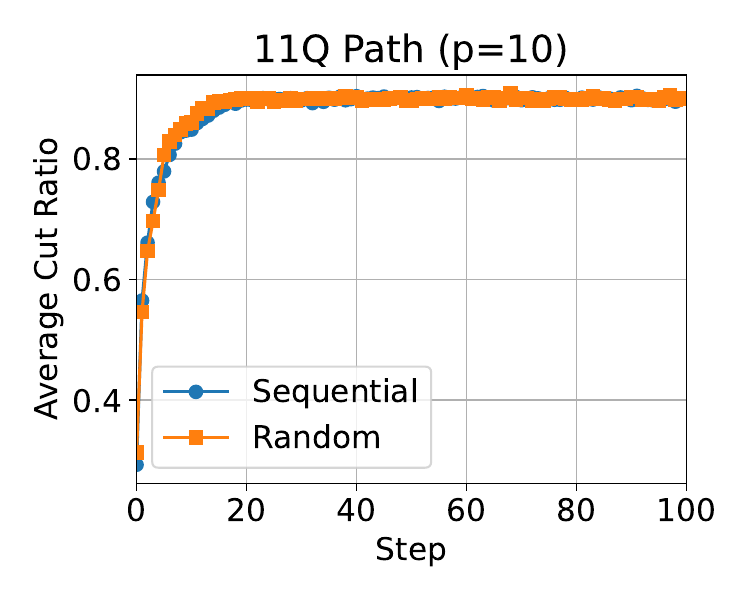}
    \end{subfigure}
    \hfill
    \begin{subfigure}[t]{0.32\textwidth}
        \centering
        \includegraphics[width=\textwidth]{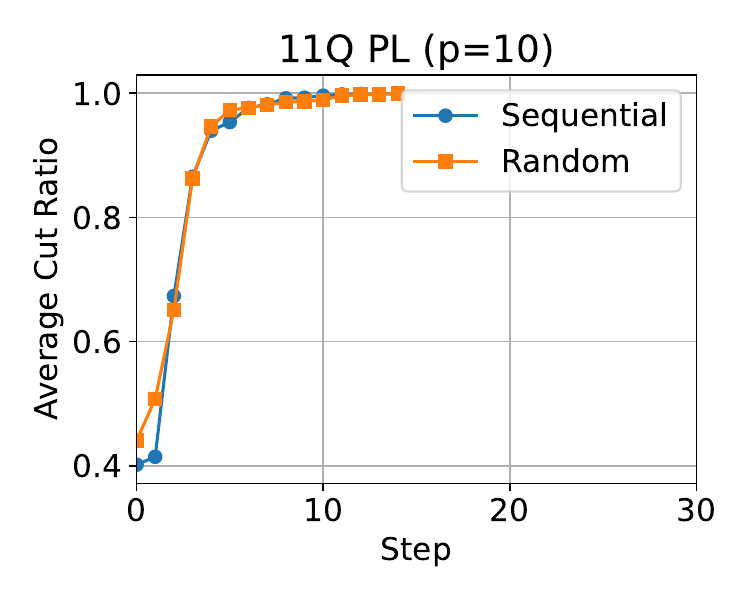}
    \end{subfigure}
    \hfill
    \begin{subfigure}[t]{0.32\textwidth}
        \centering
        \includegraphics[width=\textwidth]{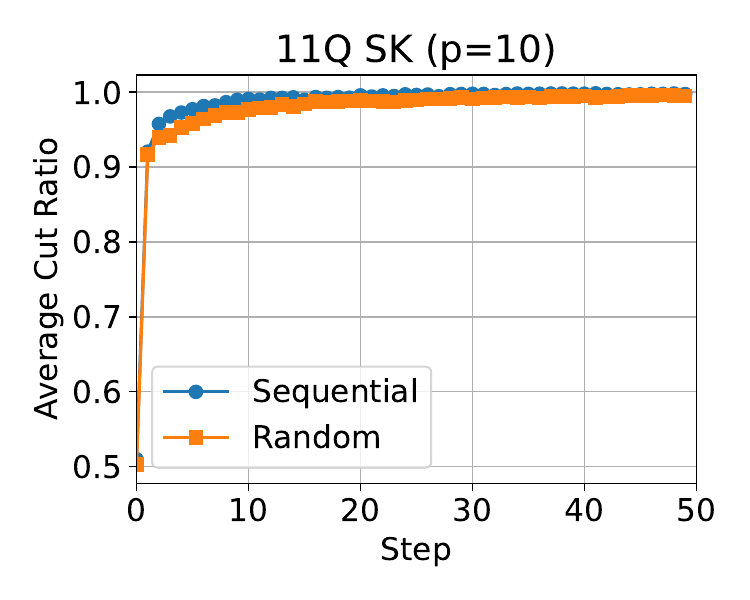}
    \end{subfigure}
    \hfill
    \caption{
        Training performance comparison of 11-qubit Orbit-QAOA circuits under the two update strategies: \textit{Sequential} vs. \textit{Random} selection.  
        Each subfigure shows the evolution of the ACR over training steps for QAOA circuits targeting different graph models (Path, Power-Law, and Sherrington-Kirkpatrick) ordered from left to right.  
        The top row corresponds to circuits with $p=5$ layers, and the bottom row corresponds to $p=10$ layers.
    }
    \label{f0}
\end{figure}

This section investigates the training efficiency of Orbit-QAOA under two different strategies for selecting which layer to update at each step.  
The goal of this experimental evaluation is to observe whether the selection order (sequential or random) of the layers to be retrained in Orbit-QAOA affects the learning efficiency.
As shown in Figure \ref{f0}, we compare the training performance of 11-qubit Orbit-QAOA circuits when layers are updated either in a \textit{sequential} or a \textit{random} order.  
Given a QAOA circuit with $k$ layers, the sequential strategy cyclically updates the layers in a fixed order—starting from layer $0$ up to layer $k{-}1$, then returning to $0$.  
In contrast, the random strategy selects one of the $k$ layers at random for update at each training step.
In this round-robin scheduling, the order of layers is randomly configured, but each cycle makes only a single visit to each layer.
Our evaluations indicate that the choice of update order has little impact on training performance.  
Across various graph models and different layer counts, both the sequential and random strategies yield comparable ACR curves.  
In MA-QAOA, each layer has sufficiently high expressibility owing to its fine-grained parameterization.  
This allows each training epoch to yield meaningful progress, even when the layer to be updated is chosen randomly.  
Consequently, the proposed Orbit approach on the MA-QAOA is expected to be relatively robust in terms of the choice of update order and does not require significantly more epochs to reach high ACR, even under randomized update schedules.

\section{Experimental Methodology}

\subsection{Evaluation Set-Up}

\subsubsection{Training Scenarios} \label{Scenarios}

We evaluate 5 QAOA circuit training scenarios: Multi-Angle QAOA (MA) \cite{herrman2022multi}, Layerwise-Multi-Angle QAOA (LMA) \cite{lavagna2024layerwise}, LMA+, round-robin Layerwise Learning (RR), and Orbit-QAOA (Orbit).  
RR and Orbit are the methods proposed in this work.  
RR represents a variant that performs only round-robin layerwise training without skipping the optimization of stabilized layers.

There is an extended variant of LMA-QAOA, eLMA-QAOA \cite{lavagna2024layerwise}, in which the circuit is first trained using LMA-QAOA and then re-optimized over all layers using the MA-QAOA strategy.  
Although eLMA-QAOA is not explicitly compared in our experimental evaluation, its performance characteristics can be reasonably inferred: the total training cost is expected to be slightly higher than LMA-QAOA, while the final expectation of the number of cuts would approximate that of MA-QAOA.

The original LMA's training policy may cause each layer to remain in each training stage longer than necessary, even after convergence.
For a fair comparison, we evaluate \texttt{LMA+} as a baseline, which incorporates the layer-skipping technique proposed in \Cref{skip}.
In other words, LMA+ immediately grafts the new layer in the next step if the ACR improvement for the current layer is lower than the set threshold.
All training scenarios except LMA-QAOA follow the unified stopping condition described in \Cref{convergence}.
The activeness threshold for all scenarios is set to $\varepsilon = 0.001$.
The performance improvements achieved by Orbit-QAOA are compared against those of MA-QAOA and \texttt{LMA+}.

We also perform evaluations on layerwise training methods for single-angle QAOA circuits in Section \ref{single}.
Single-Angle (SA) denotes the standard optimization approach in which all layers are updated. 
Layerwise Single-Angle (LSA) represents a layerwise training strategy inspired by LMA-QAOA, but applied to single-angle circuits where only one layer is optimized at a time.  
LSA+ means the unified stop condition-applied LSA.

\subsubsection{Network Models for Target Graphs}

We consider 8 network models as target graphs for the Maximum-Cut problem: Path-shaped (\texttt{Path}) \cite{jang2025mantra}, Power-Law (\texttt{PL}) \cite{ayanzadeh2023frozenqubits}, Erdős–Rényi (\texttt{ER}) \cite{erdHos2013spectral}, Barabási–Albert (\texttt{BA}) \cite{albert2002statistical}, Bianconi–Barabási (\texttt{BB}) \cite{bianconi2001bose}, Watts–Strogatz (\texttt{WS}) \cite{watts1998collective}, Sherrington-Kirkpatrick (\texttt{SK}) model \cite{basso2021quantum}, and Randomly connected (\texttt{RA}) model.
\texttt{RA} assigns a probability parameter $r$ for the existence of an edge between any pair of nodes, and ensures that the resulting graph is connected without being partitioned into disjoint subgraphs.

\subsubsection{Backend}

Evaluations of equal to or less than 20-qubit circuits were performed on a system equipped with an Intel Jasper Lake N5095 processor and 32 GB of DDR4 memory.
Evaluations of larger than 20-qubit circuits were performed on a system equipped with an AMD Threadripper PRO 3975WX processor and 1024 GB of DDR4 memory.

\subsection{QAOA Training Set-Up} \label{training_setup}

\subsubsection{Quantum Simulator} \label{qsim}

We utilize the \texttt{default.qubit} state-vector simulator provided in the PennyLane framework (Python 3.9.22, PennyLane 0.38.0) \cite{bergholm2018pennylane}.
This simulator supports both analytic (shot-free) and shot-based execution modes.
In our experiments, we employ the state-vector simulation mode with 1,024 shots to simulate finite sampling.

\subsubsection{Terminologies} \label{terminology}

Across all experimental scenarios in this work, a training step refers to a single execution of the quantum device (or simulator) followed by a single parameter update behavior.
An epoch refers to one complete round of parameter updates for the entire QAOA circuit.
For example, in the RR setting, if a QAOA circuit with $p=5$ layers executes each layer once on the quantum device (or simulator) and updates its parameters after each execution, this corresponds to $5$ training steps or $1$ epoch.

\subsubsection{Classical Optimizer}

The AdaGrad optimizer \cite{duchi2011adaptive} with a step size of 0.1 is used for all training scenarios.  
For LMA-QAOA implementations, we reinitialize the optimizer's accumulated gradient history whenever a new layer is grafted.
This is due to the reduced learning rate in the process of learning the previous layers, which may prevent the new layers from being properly trained.

\subsection{Evaluation Metrics}

\subsubsection{Approximated Cut Ratio (ACR)} \label{ACR}

ACR refers to the ratio of the expectation value of $C$ for the QAOA inference results from the simulated results to those from the real solution of max-cut \cite{choi2019tutorial, lin2024towards}.
Thus, the ACR is a real number between 0 and 1, and the closer the ACR is to 1, the closer the cost function of the results sampled from the trained QAOA circuit is to the max-cut solution of that graph.

While an ACR very close to 1 indicates that the trained QAOA circuit successfully approximates the Max-Cut solution, a suboptimal ACR after training may result from several causes.
First, if the number of QAOA layers \( p \) is too small, the variational circuit may lack sufficient expressibility to represent states that approximate the optimal Max-Cut solution, even when using the multi-angle parameterization.  
Second, when the number of parameters becomes too large, the optimization landscape may suffer from the barren plateau problem, where gradients vanish across large regions of the parameter space.  
This phenomenon makes it difficult for the classical optimizer to locate regions that yield high ACR, causing convergence to poor local optima.
Third, in scenarios where physical noise \cite{jang2023quixote} is considered, such as when deploying QAOA on real quantum hardware \cite{jang2026plutarch}, the presence of noise may further flatten the optimization landscape \cite{xue2021effects}.  
This flattening effect can suppress meaningful gradient signals and hinder the circuit’s ability to reach a high-quality solution.
Finally, when grafting a new layer after fixing previously trained parameters, if the existing fixed parameters are not sufficiently suitable for the landscape of the updated training space, updating the parameters of the new layer alone may not fully exploit the expressibility of a given QAOA circuit, which may not reach the theoretical ACR.

\subsubsection{The Number of Steps to Converge (\# Steps)} \label{num_of_steps_to_converge}

In this work, the number of steps refers to the number of training iterations required until convergence is achieved.  
For the case of original LMA-QAOA implementation \cite{lavagna2024layerwise}, as no stop condition for training is implemented, each layer is assigned a fixed number of training steps.  
For example, if a QAOA circuit consists of 5 layers and each layer is trained for 50 steps, the total number of optimization steps is 250.  
Among these, we select the earliest step index at which the ACR reaches 99.9\% of the maximum ACR observed during training as the number of steps to convergence.

Training scenarios (MA, LMA+, RR, and Orbit) except LMA adopt a unified stopping condition described in \Cref{convergence}, where training continues only while at least one active layer in the circuit surpasses the predefined improvement threshold.
Once all layers fail to exceed this threshold, indicating that their parameters have effectively converged or become frozen, the overall training process terminates.

It is important to note that a smaller number of steps does not necessarily imply that the QAOA circuit has quickly reached a good Max-Cut solution.  
The effectiveness of training should be evaluated jointly based on both the number of steps and the final ACR value.  
For instance, a small number of steps with a suboptimal ACR may indicate premature convergence to a local optimum.  
Conversely, a sufficiently long training period resulting in an ACR lower than 1.0 could be a consequence of the barren plateau phenomenon, where gradients vanish and optimization stalls despite available expressibility.

\subsubsection{Runtime Per Step (RPS)} \label{rps_metric}

This section discusses a metric that represents computational overhead per step, where the overhead includes QAOA circuit simulations and gradient update processes.
Strictly evaluating the usefulness of the QAOA algorithm would require experiments on real quantum devices, but unfortunately, access to real quantum devices that can perform the full, inexpensive, and accurate QAOA training is currently restricted.
Nevertheless, the overhead of quantum circuit simulation enables an indirect comparison of the overhead of quantum circuit execution cost.
It is important to recognize that the reduction in total simulation runtime does not necessarily translate to a proportional reduction in runtime on real quantum hardware.
In our evaluation, the comparison of runtime is primarily intended to observe the reduction in parameter-update overhead and the smaller number of optimization steps required for convergence achieved by Orbit-QAOA.
We define the runtime per step (RPS) as the total simulation runtime required to complete QAOA training divided by the number of steps to converge:

\[
\text{RPS} = \frac{\text{Total runtime (s)}}{\# \text{ Steps}}.
\]

This runtime includes quantum circuit simulations \cite{lee2025pimutation} and classical optimizer overhead for updating variational parameters.

LMA-QAOA is expected to achieve a significantly lower RPS compared to MA-QAOA.  
This is primarily because LMA-QAOA incrementally builds the QAOA circuit layer by layer, which results in shallower circuits in the early stages of training.  
Relatively shallower circuits incur lower simulation overhead, as fewer operations are executed.  
Orbit-QAOA is expected to yield a slightly lower RPS compared to MA-QAOA as well.  
Since the quantum circuit depth remains consistent between MA-QAOA and Orbit-QAOA (both use the full circuit at every step), the simulation overhead per evaluation is comparable.  
However, Orbit-QAOA updates only one layer's parameters at a time in a round-robin fashion.  
As a result, the Orbit-QAOA's gradient computation and parameter update performed by the classical optimizer are more lightweight than in MA-QAOA, where all parameters are updated simultaneously.  

\subsubsection{Gradient Improvement Per Step (GIPS)} \label{GIPS}

Gradient Improvement Per Step (GIPS) quantifies the average reduction in the cost function per training step until convergence.  
It is defined as the difference between the cost function value at the start of training and that at the convergence step, divided by the number of training steps:

\[
\text{GIPS} = \frac{C_0 - C_{\text{converged}}}{\# \text{Steps}}
\]

where \( C_0 \) denotes the cost function value at step 0, and \( C_{\text{converged}} \) refers to the value at the step where training is converged.

Since QAOA aims to minimize the cost function, GIPS would generally be a positive quantity.  
A higher GIPS indicates more substantial average progress in minimizing the cost function per training step.  
Conversely, a low GIPS can imply that there is a stagnation during the training process.
Continuous parameter updates for sufficiently trained layers may result in relatively low GIPS values.
Orbit-QAOA is expected to yield higher GIPS values than MA-QAOA and LMA-QAOA.  
By dynamically skipping stabilized layers, Orbit-QAOA concentrates optimization on those layers that remain trainable, resulting in greater cost function improvement per step.

\section{Results and Analyses} \label{result0}

\subsection{Performance Analysis of Shallow Orbit-QAOA Circuits} \label{result1}

\begin{figure}[h]
    \centering
    \begin{subfigure}[t]{0.24\textwidth}
        \centering
        \includegraphics[width=\textwidth]{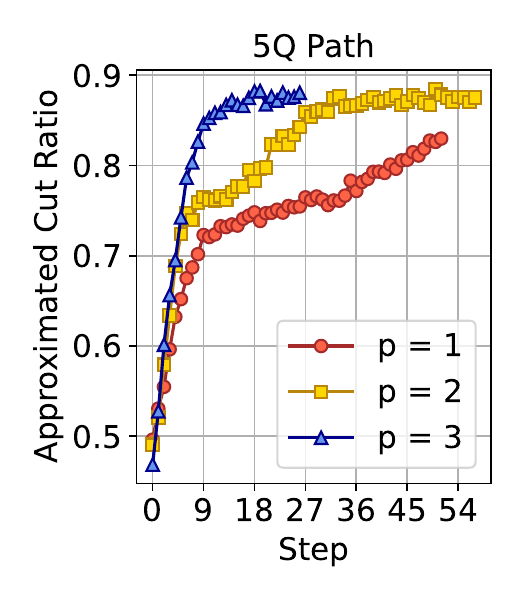}
    \end{subfigure}
    \hfill
    \begin{subfigure}[t]{0.24\textwidth}
        \centering
        \includegraphics[width=\textwidth]{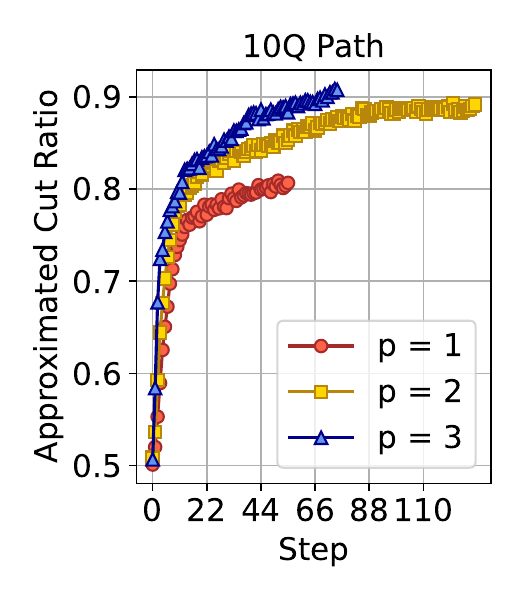}
    \end{subfigure}
    \hfill
    \begin{subfigure}[t]{0.24\textwidth}
        \centering
        \includegraphics[width=\textwidth]{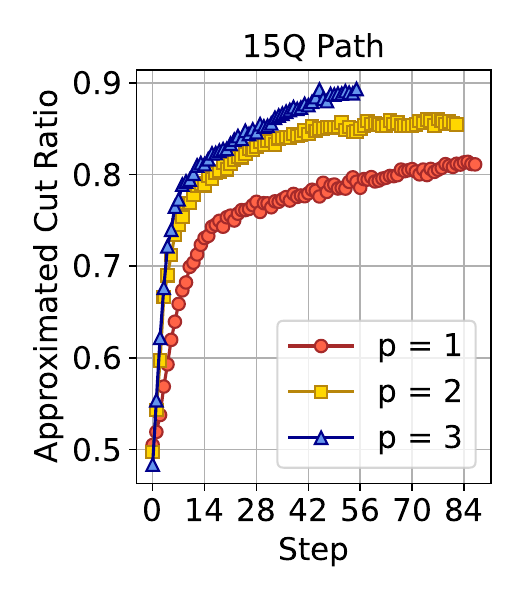}
    \end{subfigure}
    \hfill
    \begin{subfigure}[t]{0.24\textwidth}
        \centering
        \includegraphics[width=\textwidth]{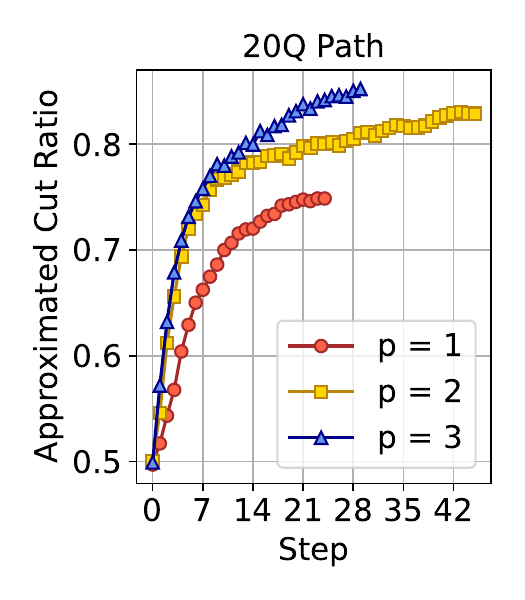}
    \end{subfigure}
    \begin{subfigure}[t]{0.24\textwidth}
        \centering
        \includegraphics[width=\textwidth]{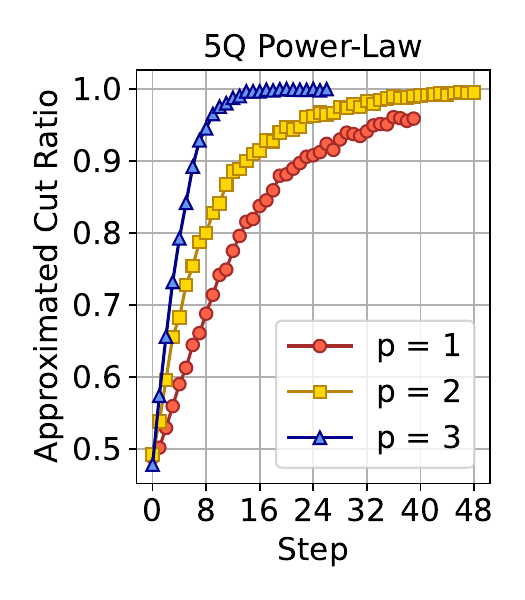}
    \end{subfigure}
    \hfill
    \begin{subfigure}[t]{0.24\textwidth}
        \centering
        \includegraphics[width=\textwidth]{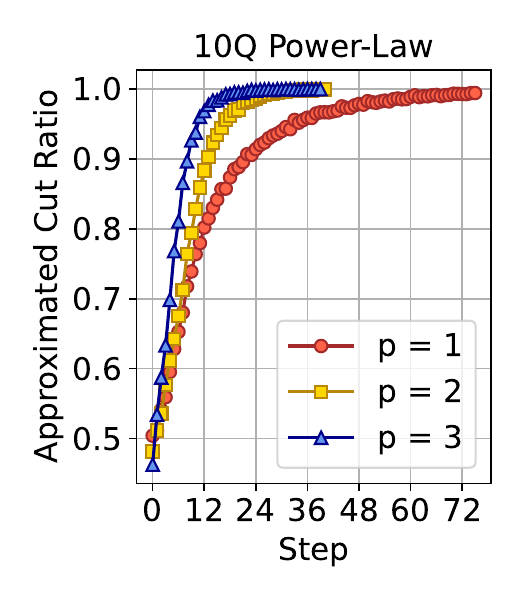}
    \end{subfigure}
    \hfill
    \begin{subfigure}[t]{0.24\textwidth}
        \centering
        \includegraphics[width=\textwidth]{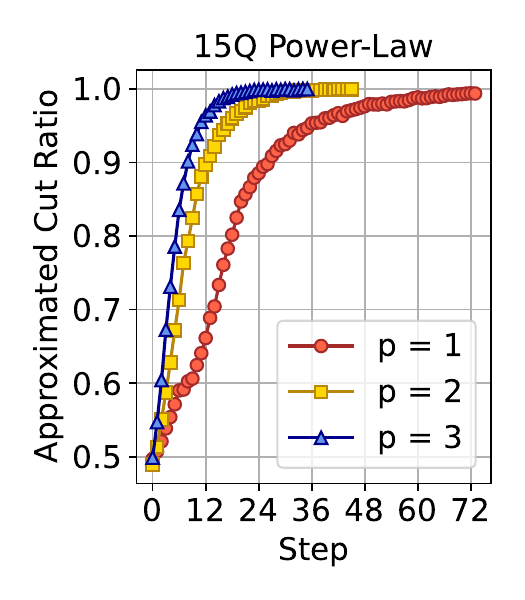}
    \end{subfigure}
    \hfill
    \begin{subfigure}[t]{0.24\textwidth}
        \centering
        \includegraphics[width=\textwidth]{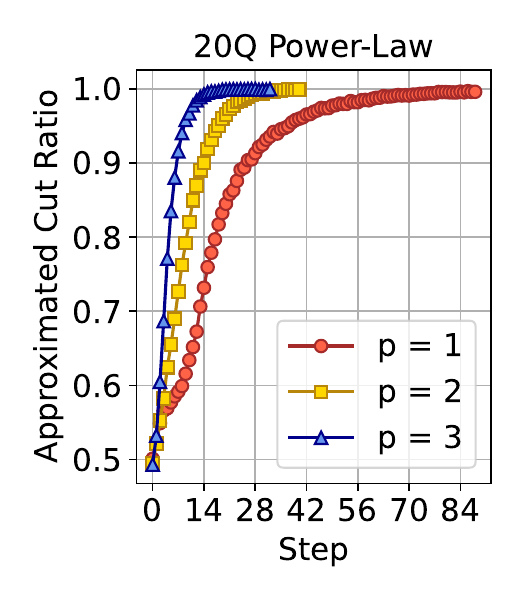}
    \end{subfigure}
        \begin{subfigure}[t]{0.24\textwidth}
        \centering
        \includegraphics[width=\textwidth]{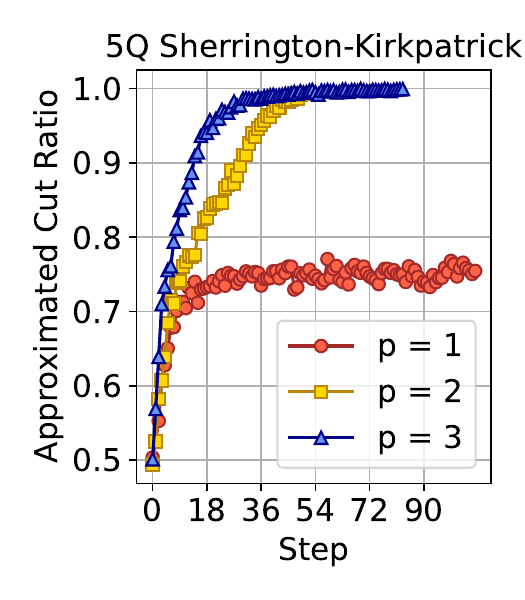}
    \end{subfigure}
    \hfill
    \begin{subfigure}[t]{0.24\textwidth}
        \centering
        \includegraphics[width=\textwidth]{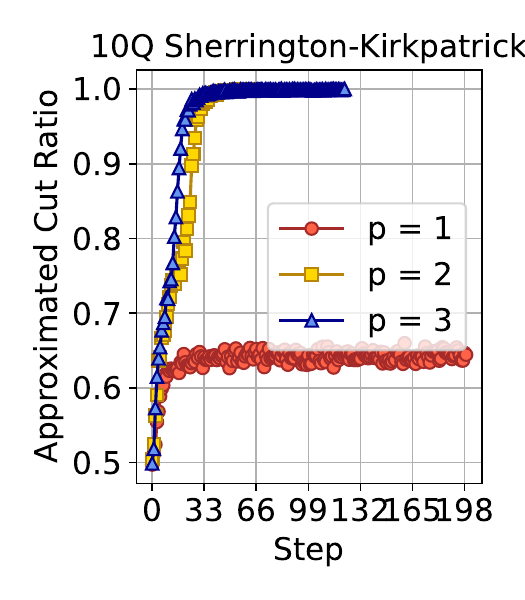}
    \end{subfigure}
    \hfill
    \begin{subfigure}[t]{0.24\textwidth}
        \centering
        \includegraphics[width=\textwidth]{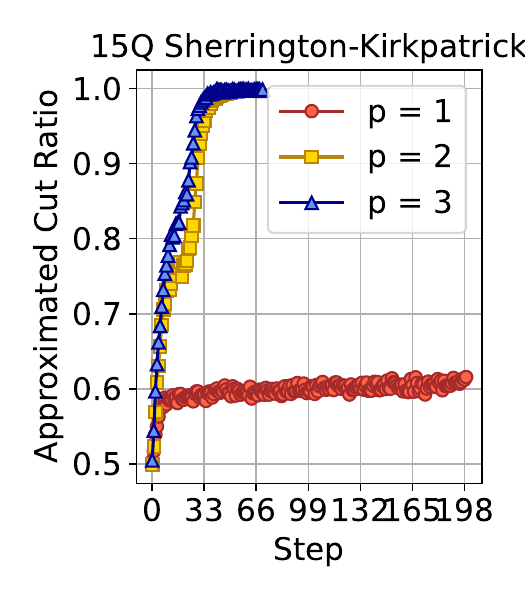}
    \end{subfigure}
    \hfill
    \begin{subfigure}[t]{0.24\textwidth}
        \centering
        \includegraphics[width=\textwidth]{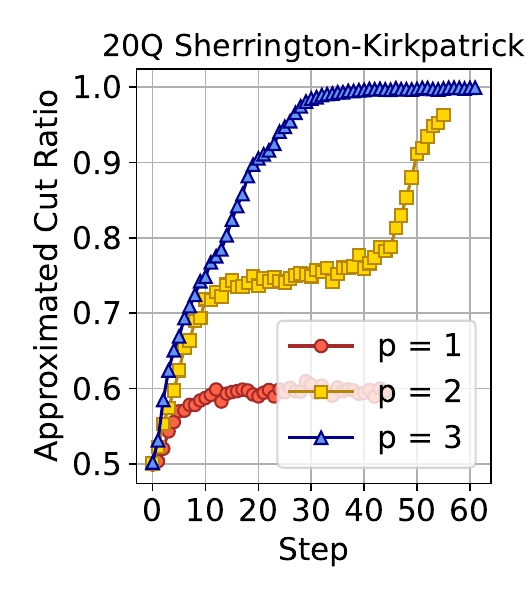}
    \end{subfigure}
    \caption{
    Evaluations of training performance of shallow Orbit-QAOA circuits across different numbers of layers (p = 1, 2, 3) and types of target graphs.
    Each subfigure shows the approximated cut ratio over training steps for graph instances with 5, 10, 15, and 20 nodes (left to right), under three target graph models: Path (top row), Power-Law (middle row), and Sherrington-Kirkpatrick (bottom row).  
    }
    \label{f1}
\end{figure}

In this section, we analyze the performance of shallow Orbit-QAOA circuits.  
As shown in Figure \ref{f1}, we evaluate QAOA circuits with $p = 1, 2, 3$ layers for target graphs constructed using three models: Path, Power-Law (PL), and Sherrington-Kirkpatrick (SK), each with 5, 10, 15, and 20 nodes.  
Table \ref{t1} provides the performances of ACR, \# Steps, and GIPS, corresponding to the training results in Figure \ref{f1}.

\begin{table}[h]
\caption{
Performance comparison of shallow Orbit-QAOA circuits across different network models and qubit sizes, with varying number of layers ($p=1,2,3$).
}
\label{t1}
\centering
\begin{tabular}{@{\extracolsep{\fill}}cc|ccc|ccc|ccc@{}}
\toprule
\multirow{2}{*}{Target} & \multirow{2}{*}{Qubits} & \multicolumn{3}{c|}{$p=1$} & \multicolumn{3}{c|}{$p=2$} & \multicolumn{3}{c}{$p=3$} \\
\cmidrule{3-11}
& & ACR & \# Steps & GIPS & ACR & \# Steps & GIPS & ACR & \# Steps & GIPS \\
\midrule
\multirow{4}{*}{Path}
& 5Q  & 0.830 & 49  & 0.03 & 0.875 & 45  & 0.03 & 0.881 & 21  & 0.07 \\
& 10Q & 0.807 & 55  & 0.05 & 0.888 & 87  & 0.04 & 0.908 & 74  & 0.05 \\
& 15Q & 0.811 & 87  & 0.05 & 0.860 & 79  & 0.06 & 0.893 & 47  & 0.12 \\
& 20Q & 0.749 & 24  & 0.20 & 0.829 & 45  & 0.14 & 0.852 & 29  & 0.23 \\
\midrule
\multirow{4}{*}{PL}
& 5Q  & 0.956 & 39  & 0.05 & 0.995 & 48  & 0.04 & 0.999 & 20  & 0.10 \\
& 10Q & 0.994 & 75  & 0.06 & 0.999 & 42  & 0.11 & 0.999 & 27  & 0.17 \\
& 15Q & 0.994 & 73  & 0.09 & 0.999 & 44  & 0.16 & 0.999 & 28  & 0.25 \\
& 20Q & 0.999 & 88  & 0.11 & 0.999 & 40  & 0.24 & 0.999 & 21  & 0.45 \\
\midrule
\multirow{4}{*}{SK}
& 5Q  & 0.755 & 107 & 0.01 & 0.994 & 51  & 0.06 & 0.995 & 47  & 0.06 \\
& 10Q & 0.654 & 184 & 0.02 & 0.999 & 66  & 0.19 & 0.999 & 48  & 0.26 \\
& 15Q & 0.616 & 199 & 0.03 & 0.997 & 49  & 0.57 & 0.999 & 51  & 0.55 \\
& 20Q & 0.596 & 44  & 0.22 & 0.963 & 57  & 0.81 & 0.999 & 53  & 0.94 \\
\midrule
\multicolumn{2}{c|}{Gmean}
& 0.801 & 71.6 & 0.06
& 0.948 & 52.8 & 0.12
& 0.959 & 35.6  & 0.18 \\
\botrule
\end{tabular}
\end{table}

\subsubsection{Impact of Increasing Layers on ACR}

As easily expected, increasing the number of layers improves the Approximate Cut Ratio (ACR).  
With $p = 2$, the geometric mean ACR reaches 0.948, which is an 18.4\% improvement over the case when $p = 1$.  
With $p = 3$, the ACR further increases to 0.959, representing a 19.7\% gain compared to $p = 1$.
In the geometric mean, GIPS is shown to be proportional to the increasing number of layers, 0.06 for p = 1, but 0.12 for p = 2 and 0.18 for p = 3.

\subsubsection{Effect of the number of Layers on Convergence Behavior}

Interestingly, the effect of increasing $p$ on convergence speed reveals a counterintuitive pattern.  
Because Orbit-QAOA updates only one layer per step, the number of trainable parameters per step is reduced compared to the MA-QAOA, which may mean that the required number of steps for convergence becomes larger.  
However, experimental results show the opposite trend: increasing the number of layers leads to faster convergence.  The average number of steps required for convergence is 26.3\% lower for $p = 2$ compared to $p = 1$, and 50.3\% lower for $p = 3$.  

This implies that the contribution of parameter updates from a single layer is sufficient to reflect meaningful ACR improvement provided by the classical optimizer.
Additionally, this improvement in convergence speed with an increasing number of layers in Orbit-QAOA may be attributed to two factors as follows.

First, deeper Orbit-QAOA circuits tend to exhibit faster ACR improvement during the early stages of training.  
This is because Orbit-QAOA updates parameters in a round-robin fashion across all layers, ensuring that each layer is trained to a similar degree during early epochs.  
As a result, the parameters across layers are guided toward similarly favorable regions in the trainable parameter space, potentially providing a better initialization for subsequent refinement.  
However, in MA-QAOA, since all parameters are updated simultaneously by the classical optimizer, this may lead to imbalanced progress.  
This imbalance may result in some layers being already sufficiently optimized while others remain undertrained, even when the overall ACR appears close to 1.

Second, Orbit-QAOA also maintains a competitive convergence rate in later training stages.  
This may be due to its layer-skipping mechanism, which removes stabilized layers from the training schedule and focuses parameter updates only on layers that remain non-converged.  
In contrast, MA-QAOA continues to update all layers uniformly, including those that may no longer contribute meaningfully to expressibility.

\subsection{Comparison with Other MA-QAOA Training Methods}  \label{result2}

\begin{figure}[h]
    \centering
    \begin{subfigure}[t]{0.32\textwidth}
        \centering
        \includegraphics[width=\textwidth]{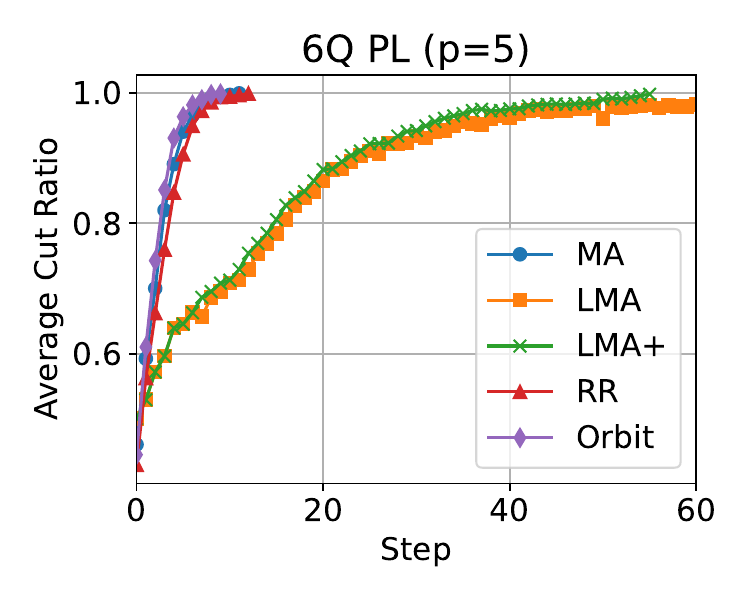}
    \end{subfigure}
    \hfill
    \begin{subfigure}[t]{0.32\textwidth}
        \centering
        \includegraphics[width=\textwidth]{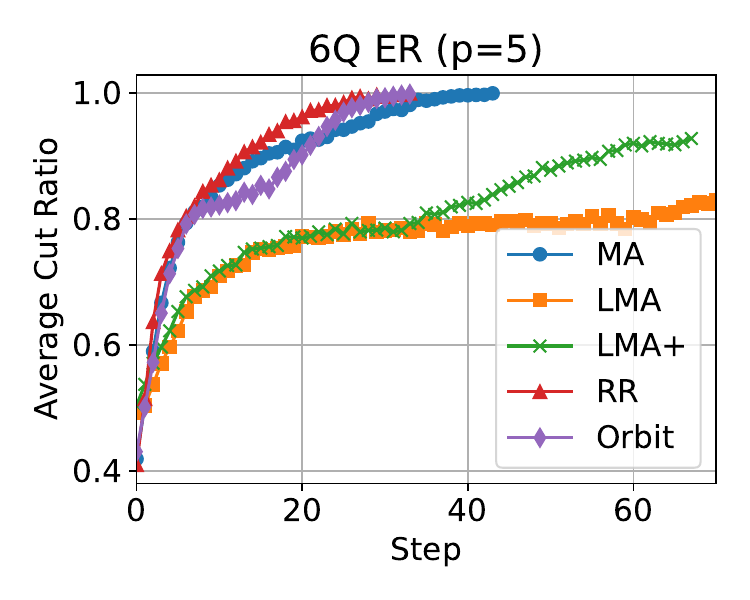}
    \end{subfigure}
    \hfill
    \begin{subfigure}[t]{0.32\textwidth}
        \centering
        \includegraphics[width=\textwidth]{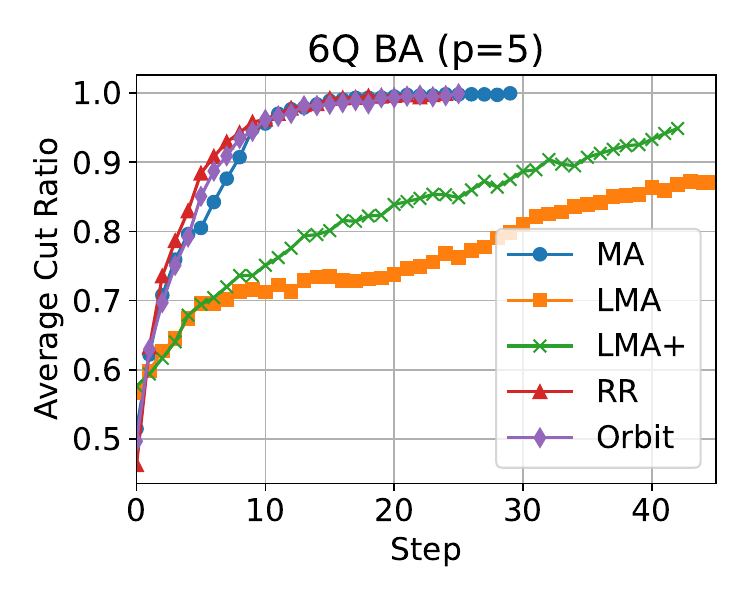}
    \end{subfigure}
    \hfill
    \begin{subfigure}[t]{0.32\textwidth}
        \centering
        \includegraphics[width=\textwidth]{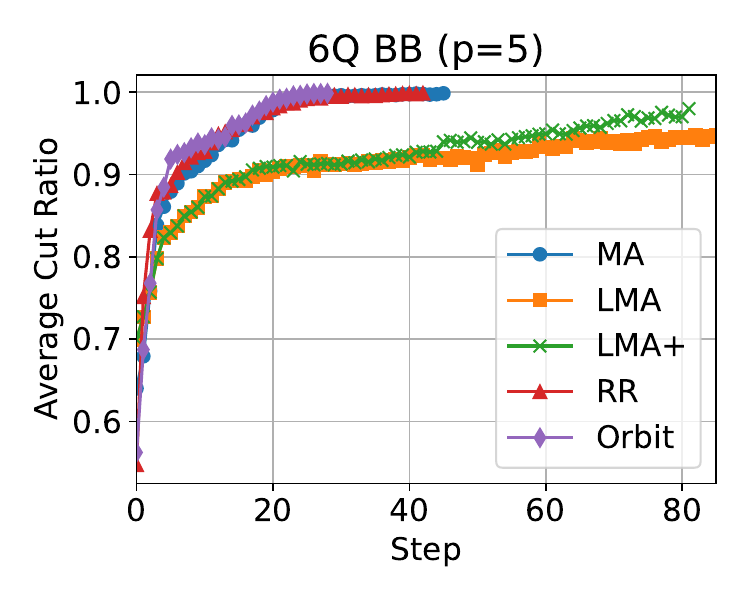}
    \end{subfigure}
    \begin{subfigure}[t]{0.32\textwidth}
        \centering
        \includegraphics[width=\textwidth]{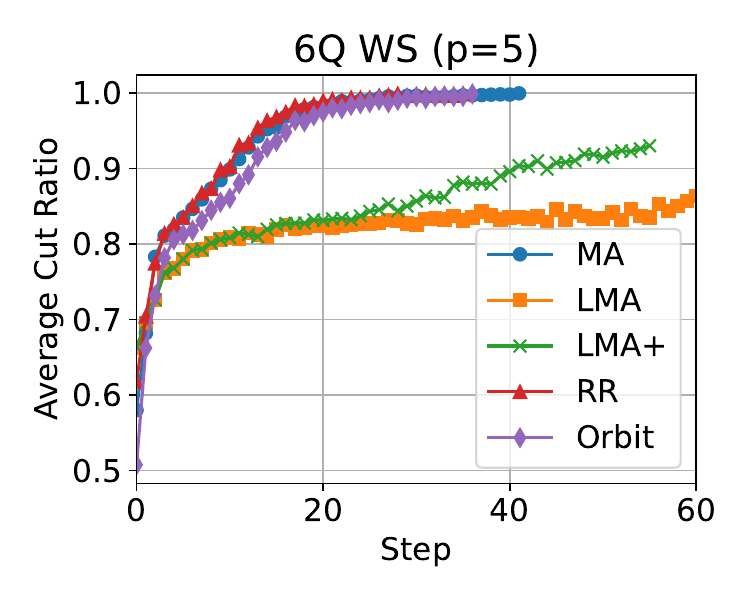}
    \end{subfigure}
    \hfill
    \begin{subfigure}[t]{0.32\textwidth}
        \centering
        \includegraphics[width=\textwidth]{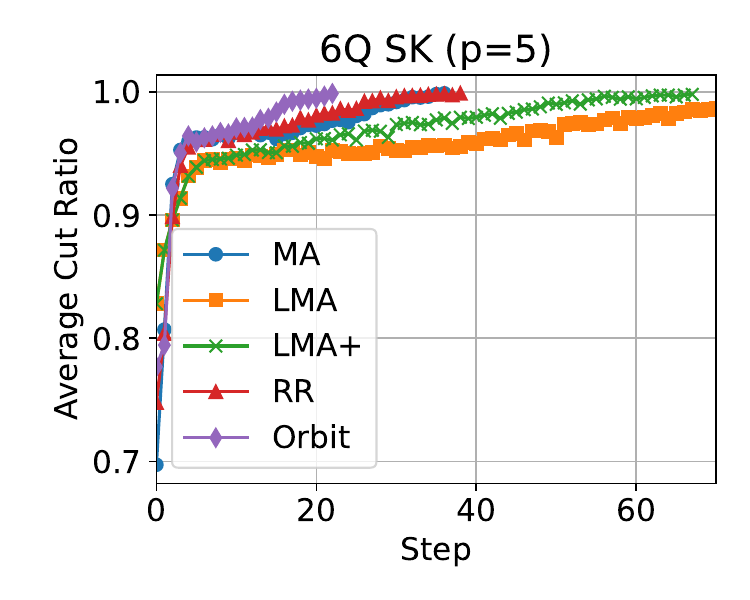}
    \end{subfigure}
    \caption{
    Comparison of 6-qubit 5-layer QAOA circuits' training processes across different optimization strategies (MA, LMA, LMA+, RR, and Orbit) on various graph models.  
    Each graph shows the evolution of the ACR over training steps for a specific network type: Power-Law (PL), Erdős–Rényi (ER), Barabási–Albert (BA), Bianconi–Barabási (BB), Watts–Strogatz (WS), and Sherrington–Kirkpatrick (SK).  
    The graphs are arranged from left to right in order of increasing average degree (i.e., number of edges divided by number of nodes).  
    }
    \label{f2}
\end{figure}

Figure \ref{f2} illustrates the training progress of 6-qubit, 5-layer QAOA circuits under different network models using five training strategies: MA, LMA, LMA+, RR, and Orbit.  
Each graph shows the evolution of the ACR as training proceeds.  
The corresponding numerical results are summarized in Table \ref{t2}.  
For only LMA-QAOA, each layer is trained for a fixed number of 50 steps (i.e., the maximum step budget is 250).
For other scenarios, a unified stop condition was applied.

\subsubsection{ACR Evaluations}

The average ACR achieved by \texttt{LMA+} is 0.971, slightly lower than 0.999, which is commonly obtained by the three non-LMA methods.  
This discrepancy can be attributed to the layer-freezing strategy employed by LMA-QAOA: once a new layer is added, the parameters of the preceding layers are fixed and no longer refined.  
As a result, earlier parameters (which are originally optimized in a shallower circuit) may not represent the best starting points in the new, expanded optimization landscape.

Indeed, we could observe a temporary degradation of the cost function improvement immediately following the introduction of a new layer in \texttt{LMA+}.  
This phenomenon suggests that the previously trained layers are no longer well aligned with the new trainable space, which may limit the optimizer’s ability to further improve the overall ACR.  
By contrast, methods such as Orbit-QAOA and RR maintain the flexibility to continually adapt all layers, enabling them to achieve consistent approximation performance to the MA-QAOA.

\begin{table}[h]
\caption{
Training results of QAOA circuits with 6 qubits and 5 layers across different graph models. 
It shows the performance of 5 training strategies (MA, LMA, LMA+, RR, and Orbit) in terms of the achieved approximated cut ratio (ACR, higher is better), total simulation runtime (in seconds, lower is better), number of training steps to convergence (lower is better), runtime per step (RPS, lower is better), and gradient improvement per step (GIPS, higher is better).  
The bold values indicate the best performance among the compared methods.
}
\label{t2}
\renewcommand{\arraystretch}{0.86} 
\centering
\begin{tabular}{cc|ccccc}
\toprule
\textbf{Graph} & \textbf{Method} & \textbf{ACR} & \textbf{Runtime (s)} & \textbf{\# Steps} & \textbf{RPS} & \textbf{GIPS} \\
\midrule
\multirow{4}{*}{6Q\_PL} & MA    & \textbf{0.999} & 22.1  & 12   & 1.84 & 0.22 \\
                        & LMA   & 0.998 & 99.8  & 173  & \textbf{0.58} & 0.01 \\
                        & LMA+  & 0.998 & 54.3  & 55   & 0.99 & 0.05 \\
                        & RR    & \textbf{0.999} & 25.6  & 13   & 1.97 & 0.22 \\
                        & Orbit & \textbf{0.999} & \textbf{19.4}  & \textbf{10}   & 1.94 & \textbf{0.28} \\
\midrule
\multirow{4}{*}{6Q\_ER} & MA    & \textbf{0.999} & 93.1  & 44   & 2.12 & 0.07 \\
                        & LMA   & 0.928 & 213.4 & 229  & \textbf{0.93} & 0.01 \\
                        & LMA+  & 0.928 & 88.1  & 67   & 1.32 & 0.03 \\
                        & RR    & \textbf{0.999} & 69.6  & \textbf{34}   & 2.05 & \textbf{0.09} \\
                        & Orbit & \textbf{0.999} & \textbf{66.6}  & \textbf{34}   & 1.96 & 0.08 \\
\midrule
\multirow{4}{*}{6Q\_BA} & MA    & \textbf{0.999} & 121.0 & 30   & 4.03 & 0.11 \\
                        & LMA   & 0.949 & 452.9 & 234  & \textbf{1.94} & 0.01 \\
                        & LMA+  & 0.949 & 122.9 & 42   & 2.93 & 0.06 \\
                        & RR    & \textbf{0.999} & 94.3  & \textbf{25}   & 3.77 & \textbf{0.15} \\
                        & Orbit & \textbf{0.999} & \textbf{85.7}  & 26   & 3.30 & 0.14 \\
\midrule
\multirow{4}{*}{6Q\_BB} & MA    & \textbf{0.999} & 239.5 & 46   & 5.21 & 0.05 \\
                        & LMA   & 0.983 & 531.0 & 243  & \textbf{2.19} & 0.01 \\
                        & LMA+  & 0.983 & 294.9 & 81   & 3.64 & 0.02 \\
                        & RR    & \textbf{0.999} & 187.6 & 43   & 4.36 & 0.07 \\
                        & Orbit & \textbf{0.999} & \textbf{132.1} & \textbf{29}   & 4.55 & \textbf{0.11} \\
\midrule
\multirow{4}{*}{6Q\_WS} & MA    & \textbf{0.999} & 280.6 & 47   & 5.97 & 0.08 \\
                        & LMA   & 0.930 & 624.6 & 248  & \textbf{2.52} & 0.01 \\
                        & LMA+  & 0.930 & 223.3 & 55   & 4.06 & 0.04 \\
                        & RR    & \textbf{0.999} & \textbf{161.4} & \textbf{31}   & 5.20 & \textbf{0.12} \\
                        & Orbit & \textbf{0.999} & 183.8 & 37   & 4.97 & \textbf{0.12} \\
\midrule
\multirow{4}{*}{6Q\_SK} & MA    & \textbf{0.999} & 349.9 & 37   & 9.46 & 0.07 \\
                        & LMA   & 0.998 & 760.6 & 245  & 3.10 & 0.01 \\
                        & LMA+  & 0.998 & 365.2 & 67   & 5.45 & 0.02 \\
                        & RR    & \textbf{0.999} & 282.6 & 39   & \textbf{7.25} & 0.06 \\
                        & Orbit & \textbf{0.999} & \textbf{164.0} & \textbf{23}   & 7.13 & \textbf{0.09} \\
\midrule
\multirow{4}{*}{Gmean}  & MA    & \textbf{0.999} & 134.3 & 32.9  & 4.08 & 0.09 \\
                        & LMA   & 0.971 & 366.8 & 227.0 & \textbf{1.62} & 0.01 \\
                        & LMA+  & 0.971 & 155.5 & 59.9  & 2.60 & 0.04 \\
                        & RR    & \textbf{0.999} & 106.3 & 28.8  & 3.69 & 0.11 \\
                        & Orbit & \textbf{0.999} & \textbf{87.2}  & \textbf{24.5}  & 3.56 & \textbf{0.12} \\
\bottomrule
\end{tabular}
\end{table}

\subsubsection{RPS Evaluations}

In terms of runtime per step (RPS), \texttt{LMA+} yields the lowest value among unified stop condition-applied methods, achieving a 1.57$\times$ reduction compared to MA-QAOA.  
This may be primarily due to its progressive training scheme, which starts from a shallow single-layer circuit and incrementally increases the depth.
RR and Orbit-QAOA achieve slightly lower RPSs compared to those of MA-QAOA (0.9$\times$ and 0.87$\times$, respectively), thanks to the reduced optimization overhead arising from layerwise parameter updates.  
Orbit-QAOA and RR producing lower average RPS values than MA-QAOA demonstrates that the proposed layerwise parameter-update scheme could reduce the classical optimizer computation overhead, despite involving the identical burdens of quantum circuit simulation workload.

\subsubsection{Evaluations for The Number of Steps to Converge}

Nevertheless, the number of steps required for convergence by \texttt{LMA+} is 1.82$\times$ higher than that of MA-QAOA, resulting in an overall simulation runtime that is 16\% longer.  
This indicates that newly grafted layers in \texttt{LMA+} could require prolonged fine-tuning processes to adjust to the fixed parameters of the earlier layers, as they struggle to improve the ACR in later stages of training.
RR and Orbit-QAOA reduce the number of steps required to reach convergence by 12.4\% and 25.4\%, respectively, compared to MA-QAOA.  
This improvement can be primarily attributed to accelerated fine-tuning during the later stages of training.  
As shown in Figure \ref{f2}, MA, RR, and Orbit exhibit similar ACR improvement trajectories during the early training phase, suggesting that the efficiency gain in RR and Orbit may stem from enhanced convergence behavior in the later epochs.
The proposed round-robin training scheme allows each layer to be optimized to a comparable extent, which can better support ACR improvements during fine-tuning.
The additional efficiency of Orbit-QAOA over RR can be explained by its layer-skipping mechanism, which avoids redundant parameter updates for already stabilized layers.

\subsubsection{GIPS Evaluations}

GIPS quantifies how much each training step contributes to improving the cost function.  
As summarized in Table \ref{t2}, Orbit-QAOA achieves the highest average GIPS of 0.12 across all benchmark graphs, followed by RR with 0.11, MA with 0.09, and \texttt{LMA+} with 0.04.  
The higher GIPS values of RR and Orbit-QAOA compared to MA-QAOA can be naturally explained by their reduced number of steps required to reach similar ACR values.  
Orbit-QAOA could further improve GIPS compared to RR by excluding stabilized layers from the training schedule, thus concentrating updates on parameters that still contribute meaningfully to optimization.

In conclusion, while \texttt{LMA+} effectively reduces the per-epoch parameter update overhead required by MA-QAOA, we observe that the early fixed layers may hinder the circuit from reaching high-precision ACRs.
In contrast, Orbit-QAOA maintains low per-epoch update costs by leveraging layer retraining and selective freezing, and also could achieve steeper ACR improvements compared to MA-QAOA.
Due to this, the following discussions and analyses would primarily focus on performance comparisons between MA-QAOA and Orbit-QAOA.

\subsection{Scalability Evaluations According to the Number of Layers} \label{result3}

\begin{table}[h]
\caption{
Comparison of ACR and the number of steps between MA-QAOA and Orbit-QAOA across various random graphs (RA) with different connectivity ratios $r$ and layer counts $p$. 
Reduced \# Steps indicates the percentage decrease in required steps by Orbit-QAOA over MA-QAOA.
}
\label{t3}
\centering
\begin{tabular}{c|c|cc|ccc}
\toprule
\multirow{2}{*}{r} & \multirow{2}{*}{$p$} & \multicolumn{2}{c|}{ACR} & \multicolumn{2}{c}{\# Steps} & \multirow{2}{*}{Reduced} \\
                        &                      & MA    & Orbit              & MA    & Orbit               &          \\
\midrule
\multirow{6}{*}{0.2} & 8  & 0.998 & 0.998 & 83 & 47 & 43.37\% \\
                     & 9  & 0.998 & 0.999 & 69 & 53 & 23.19\% \\
                     & 10 & 0.998 & 0.999 & 47 & 35 & 25.53\% \\
                     & 11 & 0.999 & 0.999 & 81 & 33 & 59.26\% \\
                     & 12 & 0.999 & 0.999 & 70 & 32 & 54.29\% \\
                     & 13 & 0.999 & 0.999 & 89 & 39 & 56.18\% \\
\midrule
\multirow{6}{*}{0.5} & 3  & 0.999 & 0.999 & 52 & 48 & 7.69\%  \\
                     & 4  & 0.999 & 0.999 & 46 & 28 & 39.13\% \\
                     & 5  & 0.999 & 0.999 & 44 & 39 & 11.36\% \\
                     & 6  & 0.999 & 0.999 & 48 & 30 & 37.50\% \\
                     & 7  & 0.999 & 0.999 & 41 & 28 & 31.71\% \\
                     & 8  & 0.999 & 0.999 & 46 & 20 & 56.52\% \\
\midrule
\multirow{6}{*}{0.8} & 3  & 0.999 & 0.999 & 80 & 56 & 30.00\% \\
                     & 4  & 0.999 & 0.999 & 53 & 47 & 11.32\% \\
                     & 5  & 0.999 & 0.999 & 52 & 39 & 25.00\% \\
                     & 6  & 0.999 & 0.999 & 47 & 42 & 10.64\% \\
                     & 7  & 0.999 & 0.999 & 37 & 32 & 13.51\% \\
                     & 8  & 0.999 & 0.999 & 37 & 31 & 16.22\% \\
\midrule
\multicolumn{2}{c|}{Gmean} & 0.999 & 0.999 & 54.56 & 36.55 & 25.52\% \\
\bottomrule
\end{tabular}
\end{table}

This section evaluates the performance of MA-QAOA and Orbit-QAOA with increasing numbers of layers across graphs with different edge connectivity ratios \( r \), as shown in Table \ref{t3}. 
The considered layer ranges in Table \ref{t3} are chosen to ensure that all circuits achieve sufficiently high ACR.  
Orbit-QAOA achieves comparable ACR values to MA-QAOA while reducing the number of required training steps, on average by 25.52\% and up to 59.26\%.
The step reduction of Orbit-QAOA becomes relatively more prominent when the target graphs have lower edge densities.

Recent studies suggest that MA-QAOA typically requires more training steps as the circuit depth increases \cite{gaidai2024performance}.  
However, our evaluation shows that this trend does not universally apply.  
Except for the specific case of MA-QAOA at \( r = 0.2 \), the number of steps required for convergence tends to decrease as the number of layers increases.
This may be attributed to the fact that increasing the number of layers improves the expressibility of the QAOA circuit, thereby accelerating the late-stage fine-tuning process required to reach high-precision ACR values near 1.
Orbit-QAOA can better exploit deeper circuits without suffering from the parameter update overhead observed in MA-QAOA, leading to both faster convergence and more scalable performance in sparse and dense target graphs.

The characteristic that deeper circuits can converge faster may imply that Orbit-QAOA may have better scalability for combinatorial optimization problems at scale. 
In other words, this scalability improvement can contribute to not only reducing the computational overhead of classical resources by the layerwise parameter updates but also reducing the computational overhead of quantum computing resources by reducing the number of epochs required for training.

\subsection{Tracking Active Layers for Orbit-QAOA Training} \label{active_freeze}

\begin{figure}[h]
    \centering
    \begin{subfigure}[t]{0.49\textwidth}
        \centering
        \includegraphics[width=\textwidth]{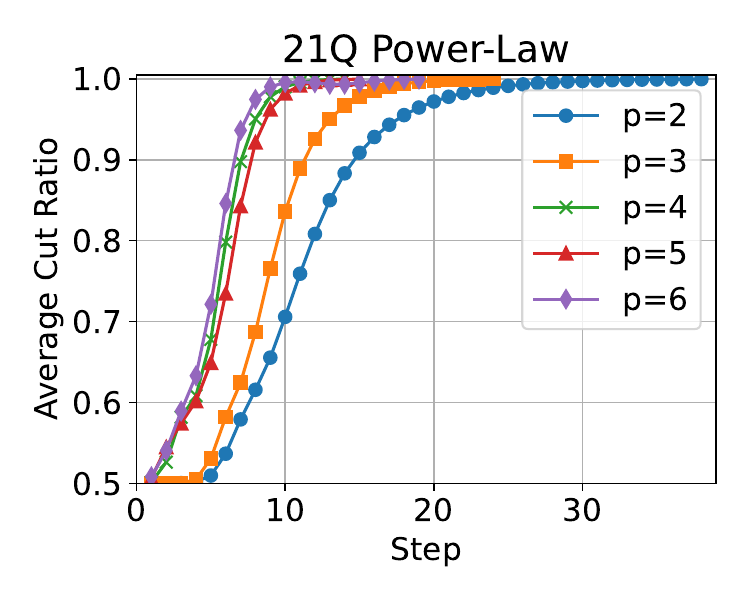}
    \end{subfigure}
    \hfill
    \begin{subfigure}[t]{0.49\textwidth}
        \centering
        \includegraphics[width=\textwidth]{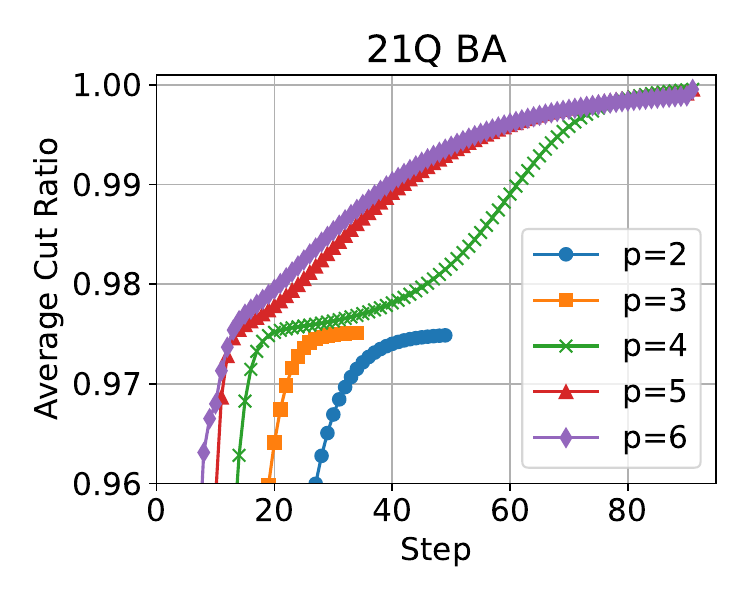}
    \end{subfigure}
    \caption{
    Training performance of  21-qubit Orbit-QAOA circuits on graphs for Power-Law (left) and Barabási–Albert (right) models under various number of layers $p$.
    }
    \label{f6}
\end{figure}

\begin{figure}[h]
        \centering
        \includegraphics[width=0.8\textwidth]{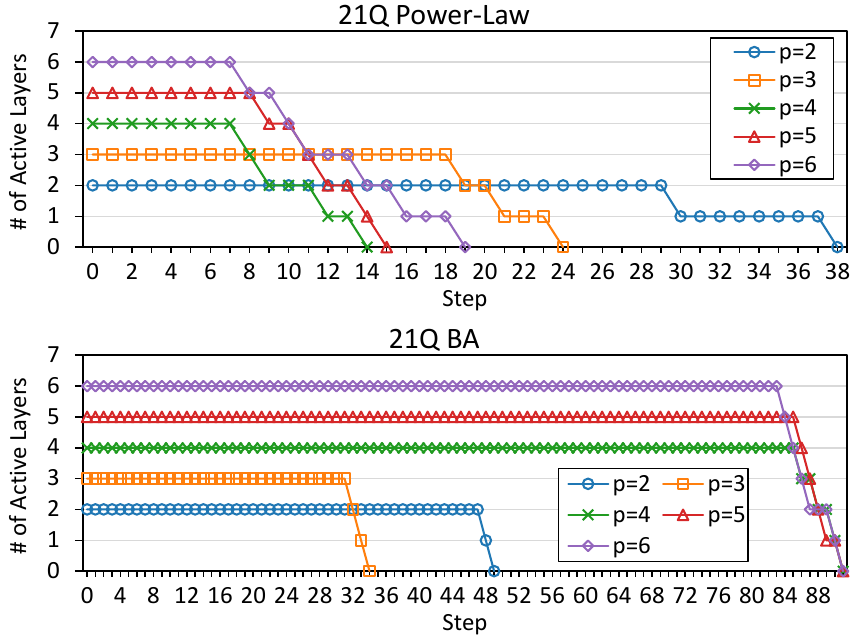}
    \hfill
    \caption{
    The number of active layers according to a different number of layers $p$ during Orbit-QAOA training for 21-qubit circuits under Power-Law (top) and Barabási–Albert (bottom) graph models.
    }
    \label{f7}
\end{figure}

This subsection analyzes how the number of active layers evolves during the Orbit-QAOA training process.
A layer is considered active when its parameters continue to be updated, whereas a frozen layer indicates that its parameters have converged below the threshold $\varepsilon = 0.001$ and have been excluded from the round-robin scheduling.
\Cref{f7} presents the number of active layers throughout the training steps of 21-qubit circuits under the Power-Law (left) and Barabási–Albert (right) graph models, each trained with different numbers of QAOA layers $p$.
In both graphs, Orbit-QAOA exhibits faster ACR improvement as $p$ increases, consistent with the trend observed in \Cref{result3}.

As shown in \Cref{f6}, for the Power-Law graph, the relatively simple and symmetric structure of the cost Hamiltonian operator allows the QAOA circuits to reach an ACR close to 1 rapidly.
Consequently, as illustrated in \Cref{f7}, the freezing process proceeds relatively quickly, meaning that after the ACR saturates near 1, the number of active layers rapidly decreases.
This suggests that, in later stages of parameter training, it is sufficient that only a few layers remain active to perform fine-tuning of parameters while the others remain frozen.

In contrast, the Barabási–Albert graph has a relatively complex cost Hamiltonian operator than the Power-Law graphs.
For the cases of $p=2$ or $p=3$ on Barabási–Albert graphs in \Cref{f6}, the limited circuit expressibility results in the training being terminated around an ACR of approximately 0.975.
Unlike the Power-Law scenario, this termination arises not from rapid convergence but from insufficient expressibility.
Due to this, all layers remain active until the end of training before freezing simultaneously.
When $p>3$, the circuit exhibits sufficient expressibility, and ACR continues to improve steadily even in later training stages.
Accordingly, as shown in \Cref{f7}, the freezing process occurs more leisurely, with most layers staying active until the late training period to collectively refine the given Max-Cut problem's solution.

Overall, these results demonstrate that the freezing behavior in Orbit-QAOA could dynamically adapt to both the structural complexity of the target graph and the circuit depth $p$.
For simple cost Hamiltonians (Power-Law graphs), the number of active layers rapidly decreases after ACR saturation, whereas for more complex graphs (Barabási–Albert), layers remain active longer for further training, which shows Orbit-QAOA's adaptive control capability.
This freezing mechanism enables Orbit-QAOA to concentrate computational effort on layers that still contribute to performance improvement, while dynamically suspending updates for converged layers.

\subsection{ACR Analysis According to Activeness Thresholds} \label{activeness}

\begin{figure}[h]
        \centering
        \includegraphics[width=0.8\textwidth]{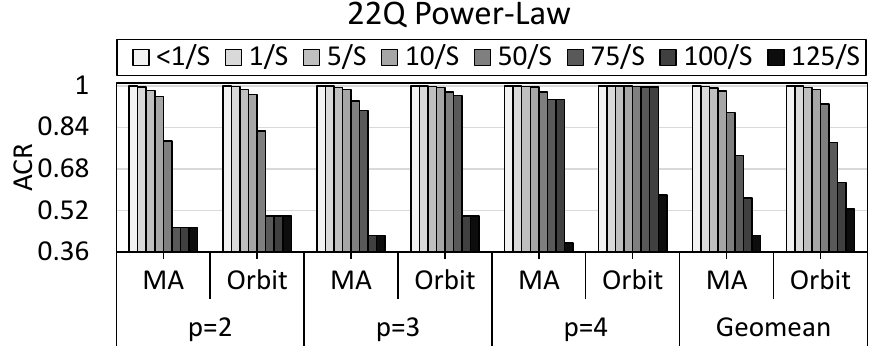}
    \hfill
    \caption{
        Comparison of the final ACR under different activeness thresholds for 22-qubit QAOA circuits of Power-Law graphs with various numbers of layers ($p =$ 2, 3, 4).
        Each bar represents the resulting ACR obtained when the activeness threshold $\varepsilon$ is scaled by the number of measurement shots $S$ = 1024, ranging from $<1/S$ to $125/S$.
    }
    \label{f9}
\end{figure}

This section investigates how varying the activeness threshold $\varepsilon$ affects the final ACR in 22-qubit QAOA circuits trained on Power-Law graphs with different layer counts ($p = 2, 3, 4$).

\subsubsection{Impact of Activeness Threshold on Final ACR}

As shown in \Cref{f9}, both MA-QAOA and Orbit-QAOA exhibit a degradation in the final ACR as the activeness threshold $\varepsilon$ increases.
For the case when $\varepsilon$ is smaller than $1/S$, the threshold is set most conservatively.
This case means that a layer is frozen only when the cost function obtained from the QAOA circuit no longer changes despite parameter updates.
In this case, most layers remain active until the theoretical ACR is reached, and both MA and Orbit achieve a final ACR of 0.9996.
When $\varepsilon$ becomes greater than or equal to $10/S$, the ACR begins to noticeably decline because several layers are prematurely frozen before full convergence is achieved.
For thresholds exceeding $75/S$, circuits with $p=2$ show that all layers freeze early, preventing meaningful training progress.
As the number of layers $p$ increases, the threshold level at which layers fail to train at all also increases, since circuits with more layers tend to require fewer steps until they approach the theoretical ACR sufficiently.
Due to this, deeper circuits with higher expressibility achieve larger average ACR improvements per step from the parameter update.
This trend is also consistent with the behavior observed in \Cref{f6} and \Cref{t3}.
In conclusion, to maintain sufficient circuit expressibility while still benefiting from effective layer freezing, setting the activeness threshold to a value slightly below $5/S$ could be recommended.

\subsubsection{Threshold Sensitivity of Orbit-QAOA versus MA-QAOA}

While both methods exhibit similar overall sensitivity of ACR degradation to threshold scaling, Orbit-QAOA consistently maintains higher ACR values than MA-QAOA across all tested thresholds.
This robustness actually stems from Orbit-QAOA’s layerwise round-robin training, which allows each layer to receive an opportunity for updates before being entirely frozen in the circuit.
In other words, Orbit has at least $p-1$ additional learning opportunities before the entire circuit freezes.
In contrast, MA-QAOA updates all layers jointly and therefore becomes more susceptible to early stagnation when the threshold is large, as the collective convergence condition may trigger global undertraining.
Thanks to this stability derived from layerwise updates, Orbit could remain more robust against ACR degradation even as the activeness threshold increases.
For instance, when the threshold is set to $125/S$, the Orbit-QAOA circuit with $p=4$ achieves the final ACR that is 24\% higher than that of MA-QAOA due to these additional learning opportunities.

\subsection{Extending Orbit to Quantum Alternating Operator Anzatz} \label{xy-mixer}

\begin{figure}[h]
    \centering
    \begin{subfigure}[t]{0.32\textwidth}
        \centering
        \includegraphics[width=\textwidth]{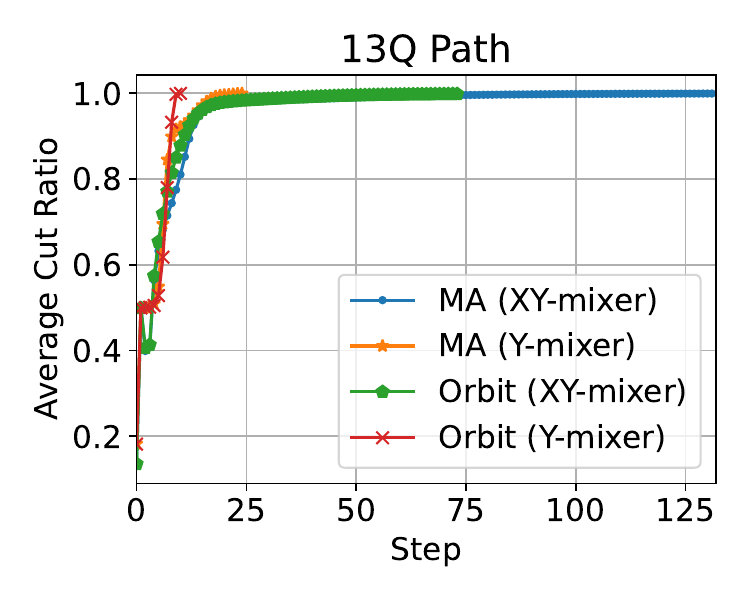}
    \end{subfigure}
    \hfill
    \begin{subfigure}[t]{0.32\textwidth}
        \centering
        \includegraphics[width=\textwidth]{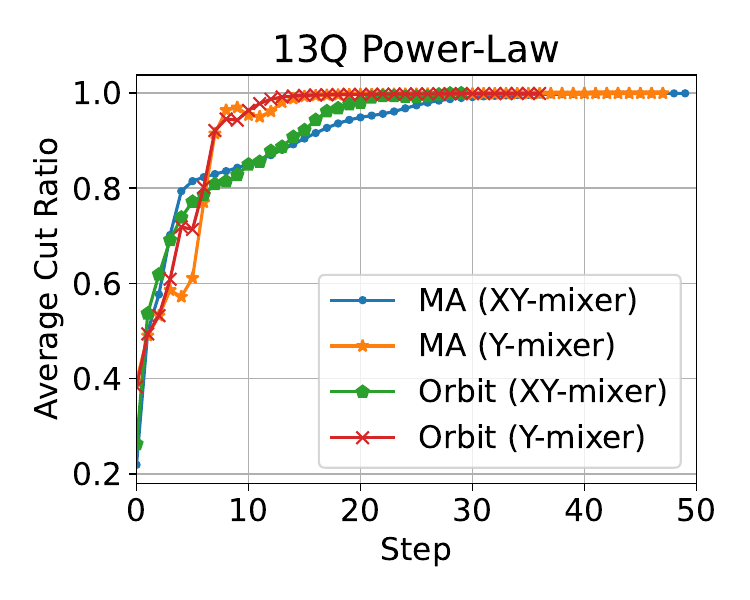}
    \end{subfigure}
    \hfill
    \begin{subfigure}[t]{0.32\textwidth}
        \centering
        \includegraphics[width=\textwidth]{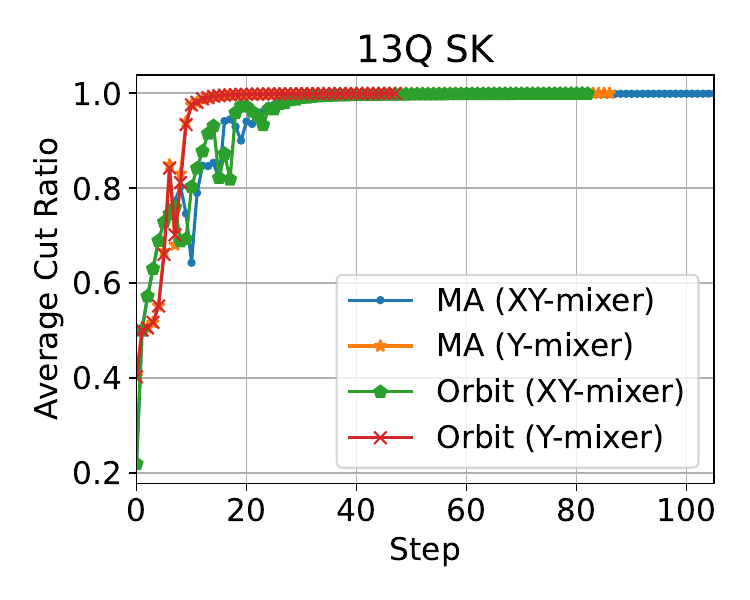}
    \end{subfigure}
    \caption{
        Training results of 13-qubit circuits with different mixer Hamiltonians under Path, Power-Law, and Sherrington–Kirkpatrick (SK) graph models.
        Each graph compares the ACR evolution between MA and Orbit-QAOA when applied to the quantum alternating operator ansatz framework \cite{hadfield2019quantum} with XY- and Y-mixers.
    }
    \label{f10}
\end{figure}

\begin{table}[h]
\caption{
Comparison of MA and Orbit performance under two quantum alternating operator ansatz frameworks with different mixer Hamiltonians (XY and Y) across Path, PL, and SK graph models. 
Each entry shows the achieved approximated cut ratio (ACR, higher is better), number of training steps to convergence (lower is better), and gradient improvement per step (GIPS, higher is better).  
The bold values indicate the best performance among the compared methods.
}
\label{t5}
\renewcommand{\arraystretch}{0.9}
\centering
\begin{tabular}{ccc|ccc}
\toprule
\textbf{Graph} & \textbf{Mixer} & \textbf{Method} & \textbf{ACR} & \textbf{\# Steps} & \textbf{GIPS} \\
\midrule
\multirow{4}{*}{Path} 
 & \multirow{2}{*}{XY-mixer} & MA    & 0.999 & 132 & 0.079 \\
 &                            & Orbit & 0.999 & 74  & 0.140 \\
 & \multirow{2}{*}{Y-mixer}  & MA    & 0.999 & 25  & 0.392 \\
 &                            & Orbit & 0.999 & 13  & 0.755 \\
\midrule
\multirow{4}{*}{PL} 
 & \multirow{2}{*}{XY-mixer} & MA    & 0.999 & 50  & 0.187 \\
 &                            & Orbit & 0.999 & 30  & 0.295 \\
 & \multirow{2}{*}{Y-mixer}  & MA    & 0.999 & 48  & 0.151 \\
 &                            & Orbit & 0.999 & 37  & 0.200 \\
\midrule
\multirow{4}{*}{SK} 
 & \multirow{2}{*}{XY-mixer} & MA    & 0.999 & 106 & 0.310 \\
 &                            & Orbit & 0.999 & 85  & 0.385 \\
 & \multirow{2}{*}{Y-mixer}  & MA    & 0.999 & 89  & 0.281 \\
 &                            & Orbit & 0.999 & 50  & 0.502 \\
\midrule
 &   \multirow{2}{*}{Geomean}    & MA    & \textbf{0.999} & 64.9 & 0.206 \\
 &                               & Orbit & \textbf{0.999} & \textbf{40.7} & \textbf{0.326} \\
\bottomrule
\end{tabular}
\end{table}

Quantum approximate optimization algorithm can be generalized into a broader framework known as the quantum alternating operator ansatz \cite{hadfield2019quantum}, which allows addressing more complex combinatorial optimization problems.  
In this section, we extend the proposed Orbit-QAOA to this generalized setting and evaluate its training performance under both XY- and Y-mixers \cite{wang2020xy, he2025non, vijendran2024expressive, jang2024recompiling}.
\Cref{f10} and \Cref{t5} show the training outcomes for 13-qubit, 5-layer QAOA circuits under three graph models of Path, Power-Law (PL), and Sherrington–Kirkpatrick (SK), where each circuit was evaluated with both XY- and Y-mixer Hamiltonians.  

\subsubsection{Performance Comparison: MA Versus Orbit}

Across all graph topologies, the final ACR remains nearly identical between MA and Orbit (0.999), confirming that Orbit’s selective layerwise update does not compromise the approximation accuracy.
However, Orbit could achieve faster convergence and greater training efficiency over the standard MA-QAOA.
On average, Orbit reduces the number of training steps by 37.3\% (from a geometric mean of 64.9 to 40.7 steps) compared to MA-QAOA.  
In terms of GIPS, Orbit improves the optimization efficiency by approximately 58.3\% (from 0.206 to 0.326 on average).

In conclusion, applying Orbit to the quantum alternating operator ansatz could extend its utility beyond the conventional quantum approximate optimization algorithm.  
This confirms that the proposed layerwise optimization and selective freezing strategy could generalize to a broader family of variational quantum algorithms. 

\subsection{Extending Orbit to Single-Angle QAOA} \label{single}

\subsubsection{Training Scenarios}

This section analyzes the training efficiency of 10-qubit single-angle QAOA circuits under different training strategies across various target graph topologies and layer counts.  
\textbf{SA} denotes the standard single-angle QAOA optimization approach. 
\textbf{LSA} represents a layerwise training strategy inspired by LMA-QAOA but applied to single-angle QAOA.  
\textbf{LSA+} is an enhanced version of LSA that incorporates the unified stopping rule, and it serves as the performance baseline.
Unlike LSA, where the number of optimization iterations per layer is predetermined, LSA+ adaptively grafts a new QAOA layer when the learning rate falls below a predefined threshold.
This adaptive mechanism could enable more efficient convergence by avoiding redundant updates on already saturated layers, thereby accelerating the overall training process compared to standard LSA.
\textbf{Orbit} corresponds to the method proposed in this work, applying the proposed Orbit strategy to single-angle QAOA by sequentially updating individual layers in a round-robin manner.

\subsubsection{Training Efficiency Analysis}

\begin{figure}[h]
    \centering
    \begin{subfigure}[t]{0.32\textwidth}
        \centering
        \includegraphics[width=\textwidth]{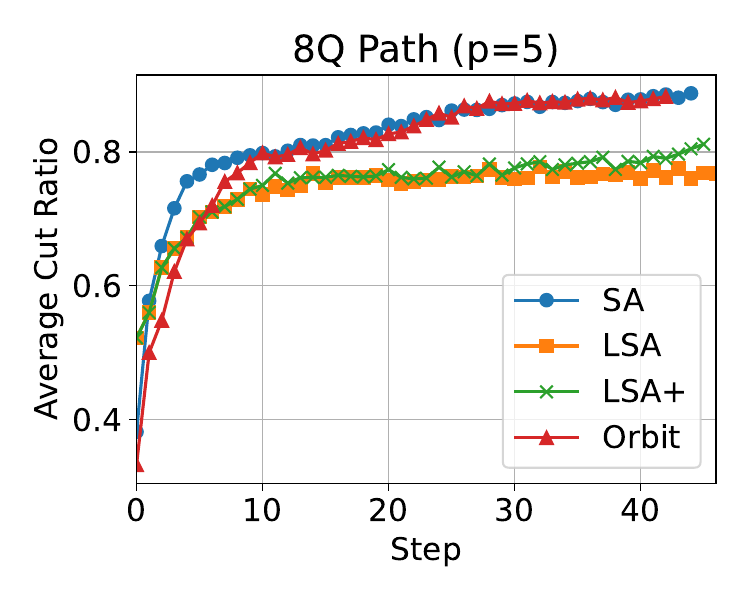}
    \end{subfigure}
    \hfill
    \begin{subfigure}[t]{0.32\textwidth}
        \centering
        \includegraphics[width=\textwidth]{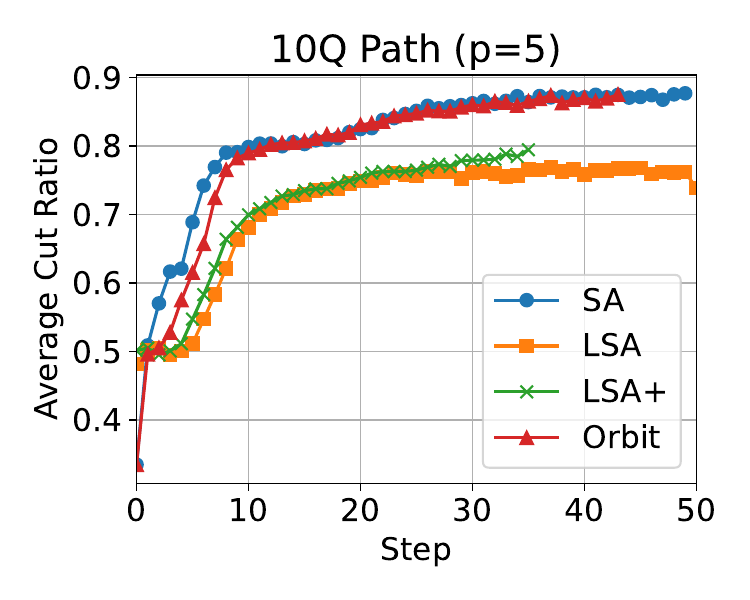}
    \end{subfigure}
    \hfill
    \begin{subfigure}[t]{0.32\textwidth}
        \centering
        \includegraphics[width=\textwidth]{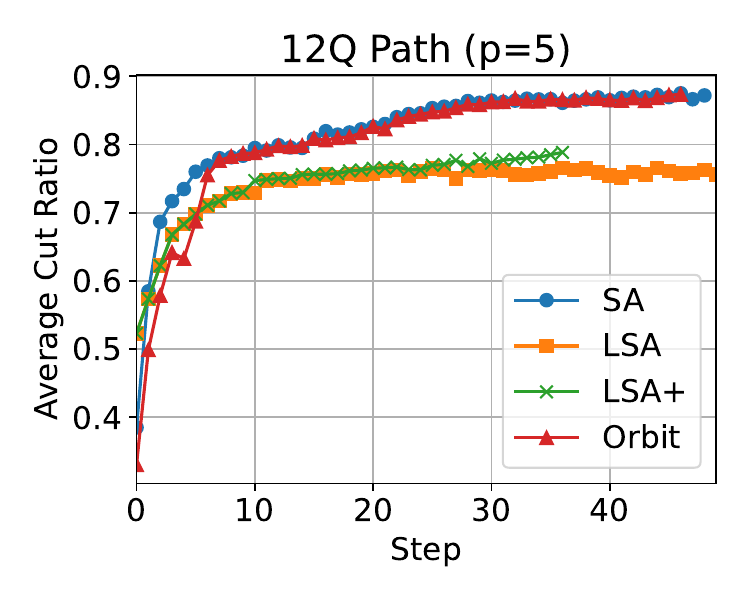}
    \end{subfigure}
    \hfill
    \begin{subfigure}[t]{0.32\textwidth}
        \centering
        \includegraphics[width=\textwidth]{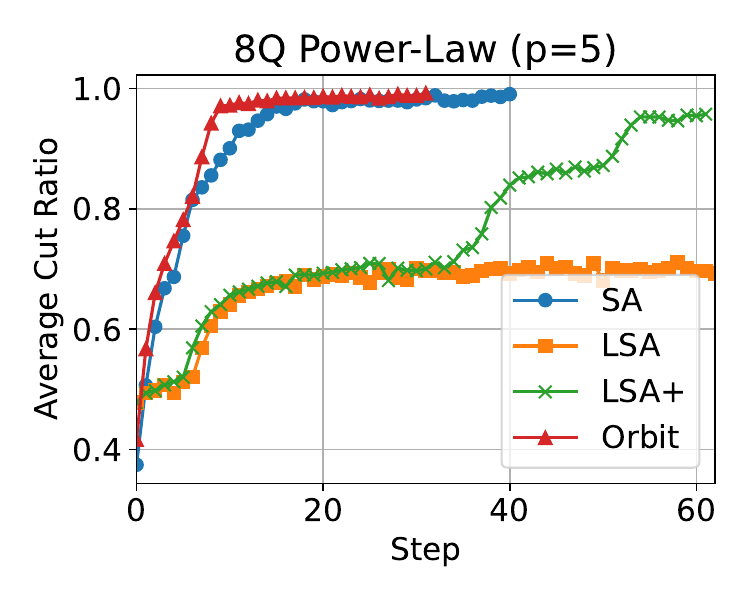}
    \end{subfigure}
    \hfill
    \begin{subfigure}[t]{0.32\textwidth}
        \centering
        \includegraphics[width=\textwidth]{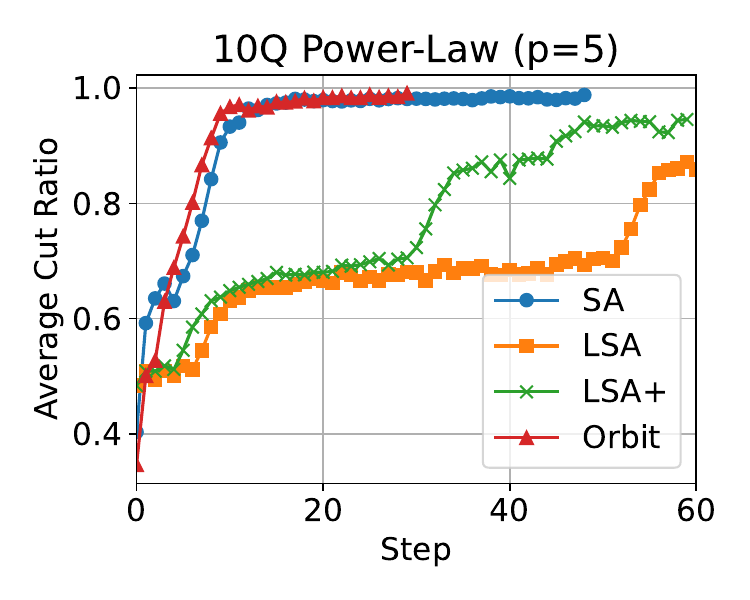}
    \end{subfigure}
    \hfill
    \begin{subfigure}[t]{0.32\textwidth}
        \centering
        \includegraphics[width=\textwidth]{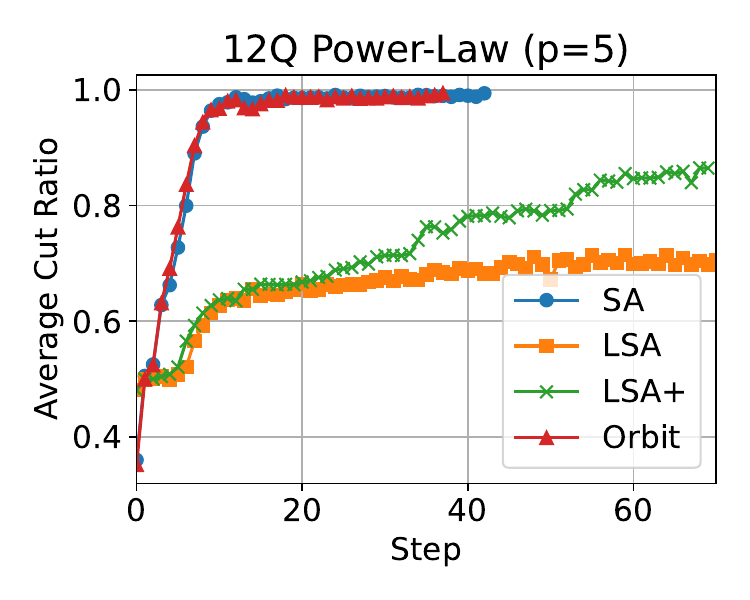}
    \end{subfigure}
    \caption{
    Comparison of training performance across different training strategies (SA, LSA, LSA+, and Orbit) for single-angle 8-qubit 5-layer QAOA circuits.  
    Each subfigure illustrates the evolution of the ACR over training steps for QAOA circuits with 8, 10, and 12 qubits (left to right), under three target graph models: Path (top row) and Power-Law (bottom row).
    }
    \label{f3}
\end{figure}

As shown in Figure \ref{f3}, the Orbit achieves a comparable final ACR to the SA baseline while requiring fewer training steps.  
Specifically, Table \ref{t4} shows that RR achieves an average ACR of 0.933 (identical to SA) while reducing the number of training steps from 45.2 to 36.2 on average, which corresponds to a 19.9\% reduction.  
The total simulation runtime is reduced by 30.4\% (from 521.6 seconds to 363.0 seconds).  
This total runtime reduction exceeds the average step reduction rate because Orbit performs parameter updates for only one layer at each step, which could decrease the classical optimization overhead compared to SA.

\subsubsection{LSA+ Versus Orbit-QAOA}

As shown in Figure \ref{f3}, the training characteristics of the LSA+ method exhibit variation depending on the target graph topology.  
For path-shaped target graph models, LSA+ may encounter difficulty achieving dramatic improvements in ACR after new layers are appended.
This suggests that the parameters in earlier layers (which are once trained and then frozen) may not function as effective warm-starts when additional layers are introduced later.  
For the Power-Law target graphs, LSA+'s SA-QAOA training trends show a clear improvement in ACR as new layers are added.  
This behavior was not observed in multi-angle QAOA evaluations, suggesting that the layer-grafting strategy by LSA+ may contribute to training speed more effectively under certain graph topologies in single-angle Hamiltonian constraints.  
This behavior may stem from the much simpler trainable parameter landscape of SA-QAOA compared to MA-QAOA.
The layerwise training and gradual layer integration strategy by LMA would be beneficial for the cases of QAOA training on certain target graph structures under SA-QAOA.

The evaluation results of LSA+ indicate that the layer-grafting strategy may have greater potential in SA-QAOA than in MA-QAOA.
Nevertheless, the average ACR of LSA+ remains 0.856, which is still 8.3\% lower than that of MA and Orbit (both 0.933).
In LSA+, the parameters of early layers that are frozen prematurely may no longer represent optimal configurations once new layers are grafted.
As a result, the LSA+'s algorithm should spend additional training steps searching new parameter combinations during the fine-tuning process in the enlarged parameter space.
Conversely, Orbit-QAOA initializes training with all layers active from the beginning, allowing global optimization across the entire circuit and enabling the achievement of higher ACR more rapidly.

\begin{table}[h]
\caption{
Comparison of ACR, number of training steps, and simulation runtime among SA, LSA, LSA+, and Orbit methods for single-angle 5-layer QAOA circuits across 2 target graph types (Path and PL) with qubit counts of 8, 10, and 12.
The bold values indicate the best performance among the compared methods.
}
\label{t4}
\centering
\begin{tabular}{cc|ccc|ccc|c}
\toprule
     &                & \multicolumn{3}{c|}{Path} & \multicolumn{3}{c|}{PL} &  \\ 
     &                & 8Q   & 10Q  & 12Q  & 8Q   & 10Q  & 12Q  & Geomean \\ 
\midrule
\multirow{4}{*}{ACR} 
     & SA             & 0.887 & 0.877 & 0.873 & 0.988 & 0.988 & 0.994 & \textbf{0.933} \\ 
     & LSA            & 0.806 & 0.797 & 0.788 & 0.953 & 0.945 & 0.863 & 0.856 \\ 
     & LSA+           & 0.806 & 0.797 & 0.788 & 0.953 & 0.945 & 0.863 & 0.856 \\ 
     & Orbit          & 0.885 & 0.875 & 0.871 & 0.992 & 0.991 & 0.994 & \textbf{0.933} \\ 
\midrule
\multirow{4}{*}{\# Steps} 
     & SA             & 45    & 50    & 45    & 40    & 49    & 43    & 45.2  \\ 
     & LSA            & 244   & 234   & 238   & 231   & 243   & 229   & 236.4 \\ 
     & LSA+           & 46    & 36    & 37    & 62    & 60    & 70    & 50.2  \\ 
     & Orbit          & 43    & 44    & 33    & 32    & 30    & 38    & \textbf{36.2}  \\ 
\midrule
\multirow{4}{*}{Runtime (s)} 
     & SA             & 237.9 & 426.3 & 1,192.7 & 302.7 & 593.9   & 925.5   & 521.6   \\ 
     & LSA            & 612.6 & 991.4 & 2,310.3 & 776.1 & 1,374.0 & 2 368.2 & 1,234.7 \\ 
     & LSA+           & 161.2 & 208.3 & 494.1   & 282.4 & 493.8   & 1,003.0 & 363.8   \\ 
     & Orbit          & 226.7 & 386.2 & 755.3   & 191.4 & 356.7   & 507.2   & \textbf{363.0}   \\ 
\bottomrule
\end{tabular}
\end{table}

\subsubsection{SA-QAOA Versus Orbit-QAOA}

Our investigations extended the Orbit-QAOA training strategy to Single-Angle QAOA (SA-QAOA), validating its effectiveness with a 20\% accelerated convergence despite reduced parameter-space complexity. 
While the proposed round-robin strategy provides training advantages in SA-QAOA as well, we would focus primarily on MA-QAOA due to its higher practical utility in the near future. 
MA-QAOA offers greater expressibility compared to SA-QAOA, enabling high approximation ratios even with fewer layers. 
Near-future quantum computers are likely to have non-negligible noise, even with quantum error correction techniques, to obtain high-precision QAOA solutions. 
In this context, MA-QAOA (despite its higher classical optimization overhead) facilitates shallow circuit designs, thereby shifting the computational burden from quantum hardware to classical resources.
Therefore, in scenarios with real quantum devices in the near future, the realistic application of the proposed round-robin layerwise update method will be relatively more advantageous when applied to MA-QAOA than to SA-QAOA.

\subsubsection{Tracing the Number of Frozen Layers in LSA+ and Orbit} \label{LSA_trace}

\begin{figure}[h]
        \centering
        \includegraphics[width=0.9\textwidth]{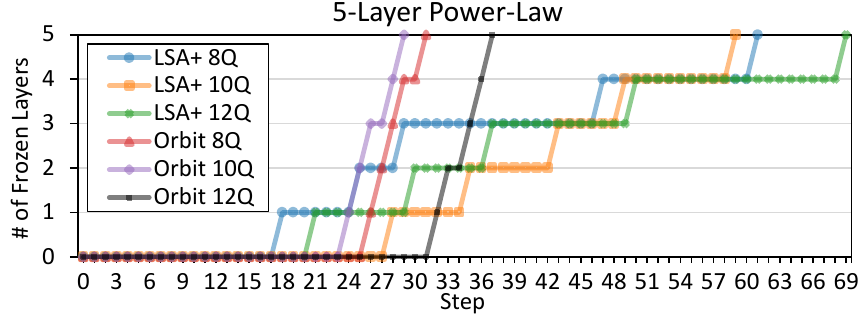}
    \hfill
    \caption{
    The number of frozen layers according to a different number of qubits under LMA+ and Orbit-QAOA training for 5-layer QAOA circuits of Power-Law graph models with 8, 10, and 12 qubits.
    }
    \label{f8}
\end{figure}

This section analyzes how the number of frozen layers evolves under different numbers of qubits for LSA+ and Orbit-QAOA.
\Cref{f8} shows the layer-freezing patterns of 5-layer QAOA circuits trained on Power-Law graphs with 8, 10, and 12 qubits.
A layer is treated as frozen once its parameter updates fall below the activeness threshold $\varepsilon = 0.001$.

Compared to LSA+, Orbit-QAOA requires fewer steps to reach convergence in all cases.
Moreover, Orbit-QAOA exhibits a pattern in which most of the layers remain active until the late stage of training and then freeze rapidly during the final fine-tuning phase.
This final stage's Orbit training pattern shows that when the QAOA circuit is reaching the theoretical ACR, all layers are driven below the activeness threshold nearly simultaneously.
By contrast, LSA+ shows a more gradual freezing trajectory.
This is because, after earlier layers are frozen, the newly grafted layers could continue to require extra training steps for additional refinement.
Relative to LSA+, Orbit-QAOA (i) consumes fewer steps for the same target graphs and (ii) triggers fast freezing of all layers at the end of training, which means QAOA circuits could be more efficiently learned within given trainable opportunities.
In LSA+, each newly appended layer prolongs training, potentially enabling slower global optimization as steps are continually spent on incremental adjustments.

\section{Related Works}

A layerwise learning strategy is a renowned approach for efficient training or compression of deep classical neural networks \cite{bengio2006greedy, fahlman1989cascade, hettinger2017forward, hinton2006fast, huang2018ltnn, teng2020layer, jangid2018handwritten}.
Layerwise learning (LL) for classical neural networks is shown to derive performance similar to training a full neural network with respect to error rates and convergence times.
Recently, layerwise learning for quantum neural networks, LL for PQC (parameterized quantum circuits), has been proposed \cite{skolik2021layerwise}.
LL for PQC learns the initial layer first, and then, once the subsequent layers are added, the previous and added layers are learned together.

Variational Quantum Eigensolver (VQE) is another representative quantum-classical hybrid algorithm that is applied in chemistry or fermionic system simulations. 
Like QAOA, which can be optimized layer by layer, VQE can also be incrementally optimized by updating individual or groups of Pauli-string terms. 
ADAPT-VQE \cite{grimsley2019adaptive} proposes an adaptive algorithm that builds the ansatz iteratively, starting from a shallow circuit and systematically grafting one operator at a time based on the specific molecular system being simulated. 
Moreover, the authors of ADAPT-VQE discuss the possibility of freezing earlier optimized parameters in later stages of training. 
This selective freezing could be conceptually aligned with techniques employed in Orbit-QAOA, which lock groups of parameters once they have sufficiently converged. 

\section{Discussion}

\subsection{Training Evaluations on Random Initial Parameters}

\begin{figure}[h]
    \centering
    \begin{subfigure}[t]{0.32\textwidth}
        \centering
        \includegraphics[width=\textwidth]{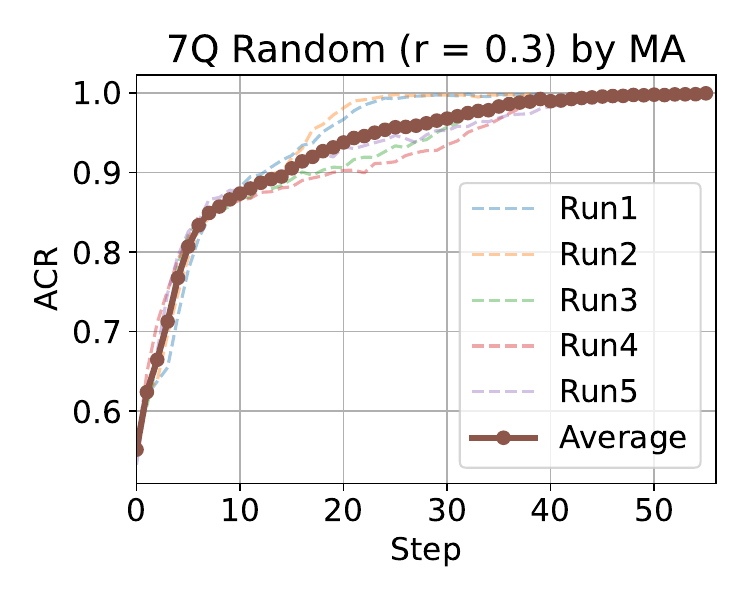}
    \end{subfigure}
    \hfill
    \begin{subfigure}[t]{0.32\textwidth}
        \centering
        \includegraphics[width=\textwidth]{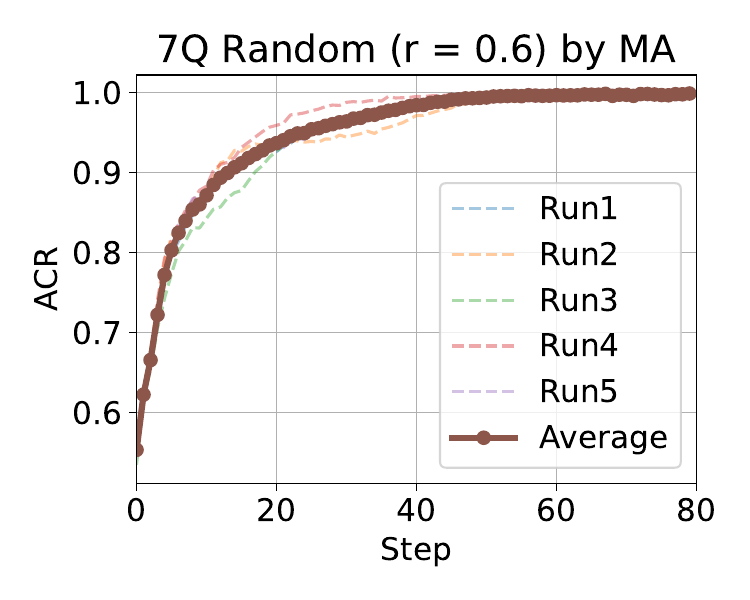}
    \end{subfigure}
    \hfill
    \begin{subfigure}[t]{0.32\textwidth}
        \centering
        \includegraphics[width=\textwidth]{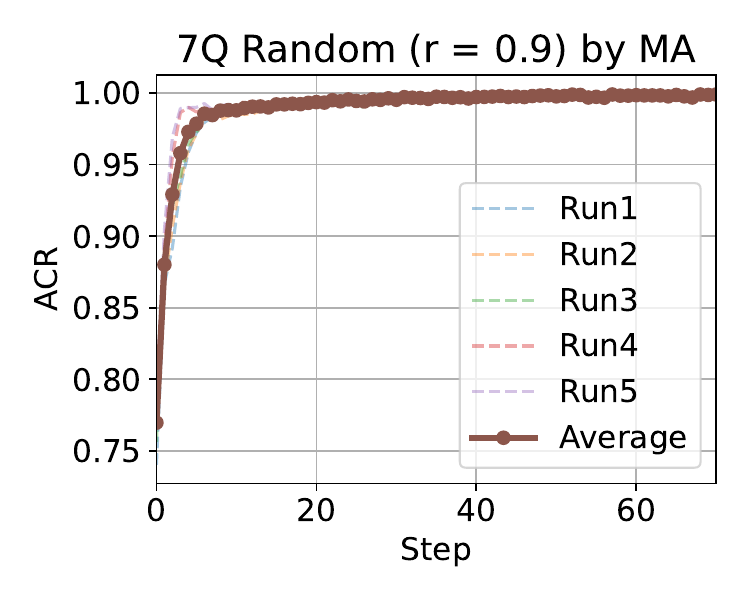}
    \end{subfigure}
    \hfill
    \begin{subfigure}[t]{0.32\textwidth}
        \centering
        \includegraphics[width=\textwidth]{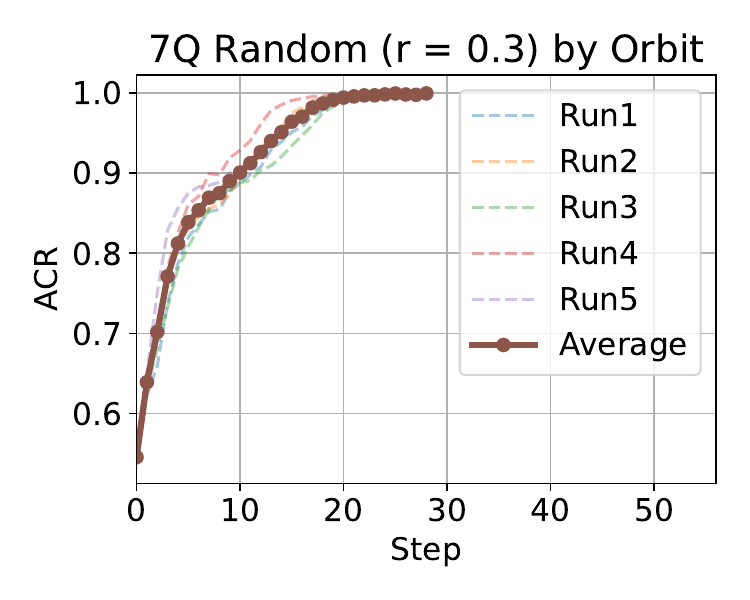}
    \end{subfigure}
    \hfill
    \begin{subfigure}[t]{0.32\textwidth}
        \centering
        \includegraphics[width=\textwidth]{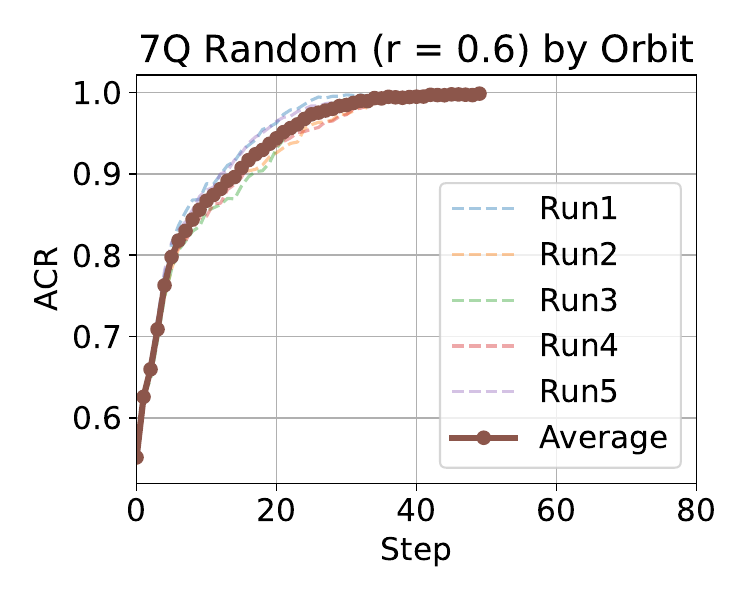}
    \end{subfigure}
    \hfill
    \begin{subfigure}[t]{0.32\textwidth}
        \centering
        \includegraphics[width=\textwidth]{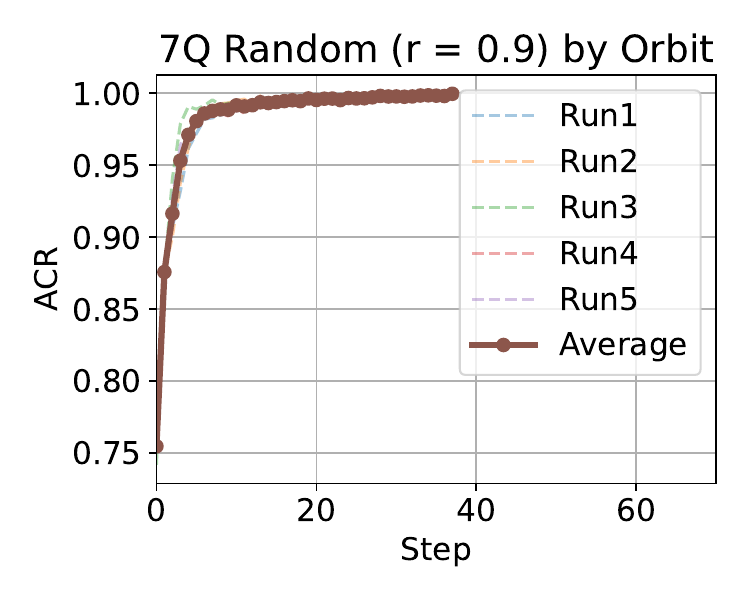}
    \end{subfigure}
    \caption{
    Comparison of training performance under five random initializations for MA-QAOA (top row) and Orbit-QAOA (bottom row) on 7-qubit 5-layer QAOA circuits with random target graphs.  
    Each subfigure illustrates the evolution of the ACRs across training steps for a specific connectivity ratio: \( r = 0.3 \), \( r = 0.6 \), and \( r = 0.9 \) (left to right).  
    MA and Orbit are plotted on different subgraphs (top and bottom, respectively) to clearly show training variations under random initializations.
    Both methods are evaluated on identical random target graph instances.
    }
    \label{f5}
\end{figure}

Figure \ref{f5} compares the training performance of MA-QAOA and Orbit-QAOA under five different random initializations on 7-qubit random graphs with varying connectivity ratios \( r \).  
Across various connectivity settings, Orbit-QAOA consistently shows faster convergence behavior compared to MA-QAOA.

The major reason behind this acceleration can vary with the connectivity of edges in the graphs.  
For sparse graphs such as \( r = 0.3 \), where circuit expressibility is relatively limited, many training steps may be required to reach high-quality solutions.  
In this case, the simplified optimization space of Orbit-QAOA (due to its layerwise parameter updates) may lead to faster progress in improving the ACR.  
On average, Orbit-QAOA requires more than 51\% fewer training steps than MA-QAOA to reach an ACR above 0.99.

In contrast, for dense graphs such as \( r = 0.9 \), where the circuit has higher expressibility, both methods quickly achieve high ACR.  
Within the first five steps, both MA and Orbit reach ACR values above 0.97, and their early-phase convergence rates appear similar.  
However, in the later training phase, Orbit-QAOA could benefit from more focused updates, refining the parameters of each layer individually.  
As a result, Orbit-QAOA requires over 35\% fewer steps to reach an ACR of 0.999, indicating more efficient fine-tuning even under random initialization.
These results suggest that Orbit-QAOA could maintain both robustness and efficiency across a range of graph densities and initial conditions, potentially.

\subsection{Appropriate Number of Layers for MA-QAOA Training}

In this section, we discuss the appropriate number of layers when applying Orbit-QAOA utilizing real quantum devices in the near future.
From the bottom line, it may depend on errors and the computational speed (i.e., operational frequency) of real quantum computing devices.
When real quantum devices have a high enough computational speed but still have a high error rate due to the relaxation \cite{jang2025balancing}, the circuit depth to perform meaningful computations is limited.
In this case, a shallow QAOA circuit should be adopted rather than trying to obtain a precise final cost function.

On the other hand, if the physical error rate of the quantum device is low enough but the operation rate is slow, it may be good to adopt a relatively deeper QAOA circuit, even if the expressibility of the circuit is already sufficient.
As discussed in Section 5.3, Orbit-QAOA tends to reduce the number of epochs required for training with deeper circuit depth.
In other words, deeper QAOA circuits can provide less overall training time, even if they have exceeded the sufficient expressibility of the trainable parameter landscape.
A potential concern for this may be the increased running time of quantum devices due to deeper circuits, but in practice, the running time of quantum devices may not be proportional to circuit depth.
In real quantum systems such as IBM's superconducting-based devices, the total runtime and energy consumption of QAOA circuits is dominated by the overhead of the qubit measurement and initialization process rather than the circuit depth \cite{jang2024recompiling}.
Thus, the increase in circuit depth in real-world device execution would not be the dominant bottleneck for execution.
In such a scenario, Orbit-QAOA with a sufficiently large number of layers is expected to quickly solve the combinatorial optimization problem with high approximation rates.
Recently reported and proposed advances in quantum error correction \cite{ai2024quantum} can make the latter case (the error of quantum devices is low enough, but their operating speed is much slower compared to classical computing) more realistic, as these techniques contribute to improving accuracy at the expense of the operating speed of quantum computing.

\section{Conclusion}

This work presents Orbit-QAOA, a round-robin layerwise optimization framework for efficiently training MA-QAOA circuits. 
Unlike conventional layer-grafting methods such as LMA-QAOA, which freeze earlier parameters during incremental training, Orbit-QAOA cyclically revisits and re-optimizes all layers iteratively until each stabilizes. 
This design allows earlier layers to adapt to changes in the trainable parameter landscape introduced by new ones.
To avoid redundant computations, Orbit-QAOA introduces a selective freezing mechanism that identifies layers with saturated convergence and skips their updates, thus improving training efficiency. 
Compared to LMA-QAOA, our method reduces training steps by up to 81.8\% and reduces approximation ratio error by up to $72\times$, while maintaining comparable approximation performance to MA-QAOA.
We further demonstrate that the round-robin training scheme is also extensible to the quantum alternating operator ansatz and single-angle QAOA, and that the granularity of layerwise updates could outperform that of sub-layer updates.
This work can contribute toward scalable quantum-classical optimization workflows, particularly relevant in settings where high-performance classical resources could play a central role in supporting quantum applications.

\section*{Acknowledgements}

The authors extend gratitude to the reviewers and editors for their constructive comments.

\section*{Author Contributions}

\noindent E. J. conceived the project idea, conducted numerical experiments, and wrote a draft manuscript.
Z. C. illustrated the idea's overview and the results of the experiments.
D. H. edited the idea methodology section.
S. C. investigated and analyzed related works.
Y. L. edited the background section.
J. K. edited the motivation section.
E. Z. Z. analyzed the results of the numerical experiments.
Y. H. edited the introduction and observation sections and directed the writing flow for the structure of numerical experiments.
W. W. R. supervised the research.

\section*{Funding}

\noindent  
This work was supported by the National Research Foundation of Korea (NRF) under the project ``Creation of the Quantum Information Science R\&D Ecosystem Based on Human Resource'' (Grant RS-2023-00303229), and by the United States Department of Energy (DOE) under Award DE-SC0025563.
This research was also supported by the education and training program of the Quantum Information
Research Support Center, funded via the NRF by the Ministry of Science and ICT (MSIT) of the Korean government (Grant RS-2023-NR057243). 

\section*{Additional Information}

\subsection*{Publisher's Note}

Springer Nature remains neutral with regard to jurisdictional claims in published maps and institutional affiliations.

\subsection*{Data Availability} 

\noindent No datasets were generated or analyzed during the
current study.

\subsection*{Code Availability} 

\noindent The code is available at a reasonable request from the
corresponding author.

\section*{Declarations}

\noindent \textbf{Competing Interests}: The authors declare no competing interests.


\bibliography{sn-bibliography}

@article{farhi2014quantum,
  title={A quantum approximate optimization algorithm},
  author={Farhi, Edward and Goldstone, Jeffrey and Gutmann, Sam},
  journal={arXiv preprint arXiv:1411.4028},
  year={2014}
}

@inproceedings{srivastava2025improved,
  title={Improved Performance of Multi-Angle Quantum Approximate Optimization Algorithm (ma-QAOA) Compared to QAOA On Simulation and Experimental Hardware Platforms},
  author={Srivastava, Vandit and Rohith, P and Singhal, Sanyam and Bhowmik, Debanjan},
  booktitle={2025 17th International Conference on COMmunication Systems and NETworks (COMSNETS)},
  pages={1130--1135},
  year={2025},
  organization={IEEE}
}

@inproceedings{shi2022multiangle,
  title={Multiangle qaoa does not always need all its angles},
  author={Shi, Kaiyan and Herrman, Rebekah and Shaydulin, Ruslan and Chakrabarti, Shouvanik and Pistoia, Marco and Larson, Jeffrey},
  booktitle={2022 IEEE/ACM 7th Symposium on Edge Computing (SEC)},
  pages={414--419},
  year={2022},
  organization={IEEE}
}

@article{gaidai2024performance,
  title={Performance analysis of multi-angle QAOA for p> 1},
  author={Gaidai, Igor and Herrman, Rebekah},
  journal={Scientific Reports},
  volume={14},
  number={1},
  pages={18911},
  year={2024},
  publisher={Nature Publishing Group UK London}
}

@inproceedings{lavagna2024layerwise,
  title={A Layerwise-Multi-Angle Approach to Fine-Tuning the Quantum Approximate Optimization Algorithm},
  author={Lavagna, Leonardo and Ceschini, Andrea and Rosato, Antonello and Panella, Massimo},
  booktitle={2024 International Joint Conference on Neural Networks (IJCNN)},
  pages={1--8},
  year={2024},
  organization={IEEE}
}

@article{herrman2022multi,
  title={Multi-angle quantum approximate optimization algorithm},
  author={Herrman, Rebekah and Lotshaw, Phillip C and Ostrowski, James and Humble, Travis S and Siopsis, George},
  journal={Scientific Reports},
  volume={12},
  number={1},
  pages={6781},
  year={2022},
  publisher={Nature Publishing Group UK London}
}

@inproceedings{jang2024recompiling,
  title={Recompiling QAOA Circuits on Various Rotational Directions},
  author={Jang, Enhyeok and Ha, Dongho and Choi, Seungwoo and Kim, Youngmin and Kwon, Jaewon and Lee, Yongju and Ahn, Sungwoo and Kim, Hyungseok and Ro, Won Woo},
  booktitle={Proceedings of the 2024 International Conference on Parallel Architectures and Compilation Techniques},
  pages={309--324},
  year={2024}
}

@inproceedings{acampora2024application,
  title={Application of Quantum Genetic Algorithms to Connected and Electric Vehicles Energy Consumption Optimization},
  author={Acampora, Giovanni and Chiatto, Angela and De Luca, Stefano and Di Pace, Roberta and Fiori, Chiara and Landolfi, Enrico and Massa, Alfredo and Schiattarella, Roberto and Vitiello, Autilia},
  booktitle={2024 IEEE 8th Forum on Research and Technologies for Society and Industry Innovation (RTSI)},
  pages={530--535},
  year={2024},
  organization={IEEE}
}

@article{tennie2025quantum,
  title={Quantum computing for nonlinear differential equations and turbulence},
  author={Tennie, Felix and Laizet, Sylvain and Lloyd, Seth and Magri, Luca},
  journal={Nature Reviews Physics},
  pages={1--11},
  year={2025},
  publisher={Nature Publishing Group UK London}
}

@article{majumder2024variational,
  title={Variational measurement-based quantum computation for generative modeling},
  author={Majumder, Arunava and Krumm, Marius and Radkohl, Tina and Fiderer, Lukas J and Nautrup, Hendrik Poulsen and Jerbi, Sofiene and Briegel, Hans J},
  journal={Physical Review A},
  volume={110},
  number={6},
  pages={062616},
  year={2024},
  publisher={APS}
}

@article{vo2025q,
  title={Q-MARL: A quantum-inspired algorithm using neural message passing for large-scale multi-agent reinforcement learning},
  author={Vo, Kha and Lin, Chin-Teng},
  journal={arXiv preprint arXiv:2503.07397},
  year={2025}
}

@article{rieffel2024assessing,
  title={Assessing and advancing the potential of quantum computing: A NASA case study},
  author={Rieffel, Eleanor G and Asanjan, Ata Akbari and Alam, M Sohaib and Anand, Namit and Neira, David E Bernal and Block, Sophie and Brady, Lucas T and Cotton, Steve and Izquierdo, Zoe Gonzalez and Grabbe, Shon and others},
  journal={Future Generation Computer Systems},
  volume={160},
  pages={598--618},
  year={2024},
  publisher={Elsevier}
}

@article{incudini2024automatic,
  title={Automatic and effective discovery of quantum kernels},
  author={Incudini, Massimiliano and Bosco, Daniele Lizzio and Martini, Francesco and Grossi, Michele and Serra, Giuseppe and Di Pierro, Alessandra},
  journal={IEEE Transactions on Emerging Topics in Computational Intelligence},
  year={2024},
  publisher={IEEE}
}

@inproceedings{ovide2024scaling,
  title={Scaling and assigning resources on ion trap QCCD architectures},
  author={Ovide, Anabel and Cuomo, Daniele and Almudever, Carmen G},
  booktitle={2024 IEEE International Conference on Quantum Computing and Engineering (QCE)},
  volume={1},
  pages={959--970},
  year={2024},
  organization={IEEE}
}

@article{kalis2023hybrid,
  title={A hybrid quantum-classical approach for inference on restricted Boltzmann machines},
  author={K{\=a}lis, M{\=a}rti{\c{n}}{\v{s}} and Loc{\=a}ns, Andris and {\v{S}}ikovs, Rolands and Naseri, Hassan and Ambainis, Andris},
  journal={Quantum Machine Intelligence},
  volume={5},
  number={2},
  pages={44},
  year={2023},
  publisher={Springer}
}

@article{duchi2011adaptive,
  title={Adaptive subgradient methods for online learning and stochastic optimization.},
  author={Duchi, John and Hazan, Elad and Singer, Yoram},
  journal={Journal of machine learning research},
  volume={12},
  number={7},
  year={2011}
}

@article{bengio2006greedy,
  title={Greedy layer-wise training of deep networks},
  author={Bengio, Yoshua and Lamblin, Pascal and Popovici, Dan and Larochelle, Hugo},
  journal={Advances in neural information processing systems},
  volume={19},
  year={2006}
}

@article{fahlman1989cascade,
  title={The cascade-correlation learning architecture},
  author={Fahlman, Scott and Lebiere, Christian},
  journal={Advances in neural information processing systems},
  volume={2},
  year={1989}
}

@article{hettinger2017forward,
  title={Forward thinking: Building and training neural networks one layer at a time},
  author={Hettinger, Chris and Christensen, Tanner and Ehlert, Ben and Humpherys, Jeffrey and Jarvis, Tyler and Wade, Sean},
  journal={arXiv preprint arXiv:1706.02480},
  year={2017}
}

@article{hinton2006fast,
  title={A fast learning algorithm for deep belief nets},
  author={Hinton, Geoffrey E and Osindero, Simon and Teh, Yee-Whye},
  journal={Neural computation},
  volume={18},
  number={7},
  pages={1527--1554},
  year={2006},
  publisher={MIT Press One Rogers Street, Cambridge, MA 02142-1209, USA journals-info~…}
}

@article{skolik2021layerwise,
  title={Layerwise learning for quantum neural networks},
  author={Skolik, Andrea and McClean, Jarrod R and Mohseni, Masoud and Van Der Smagt, Patrick and Leib, Martin},
  journal={Quantum Machine Intelligence},
  volume={3},
  pages={1--11},
  year={2021},
  publisher={Springer}
}

@article{huang2018ltnn,
  title={LTNN: A layerwise tensorized compression of multilayer neural network},
  author={Huang, Hantao and Yu, Hao},
  journal={IEEE transactions on neural networks and learning systems},
  volume={30},
  number={5},
  pages={1497--1511},
  year={2018},
  publisher={IEEE}
}

@article{teng2020layer,
  title={The layer-wise training convolutional neural networks using local loss for sensor-based human activity recognition},
  author={Teng, Qi and Wang, Kun and Zhang, Lei and He, Jun},
  journal={IEEE Sensors Journal},
  volume={20},
  number={13},
  pages={7265--7274},
  year={2020},
  publisher={IEEE}
}

@article{jangid2018handwritten,
  title={Handwritten devanagari character recognition using layer-wise training of deep convolutional neural networks and adaptive gradient methods},
  author={Jangid, Mahesh and Srivastava, Sumit},
  journal={journal of imaging},
  volume={4},
  number={2},
  pages={41},
  year={2018},
  publisher={MDPI}
}

@article{jang2025mantra,
  title={Mantra: Rewriting Quantum Programs to Minimize Trap-Movements for Zoned Rydberg Atom Arrays},
  author={Jang, Enhyeok and Kim, Youngmin and Kim, Hyungseok and Choi, Seungwoo and Huang, Yipeng and Ro, Won Woo},
  journal={arXiv preprint arXiv:2503.02272},
  year={2025}
}

@inproceedings{ayanzadeh2023frozenqubits,
  title={Frozenqubits: Boosting fidelity of qaoa by skipping hotspot nodes},
  author={Ayanzadeh, Ramin and Alavisamani, Narges and Das, Poulami and Qureshi, Moinuddin},
  booktitle={Proceedings of the 28th ACM International Conference on Architectural Support for Programming Languages and Operating Systems, Volume 2},
  pages={311--324},
  year={2023}
}

@article{basso2021quantum,
  title={The quantum approximate optimization algorithm at high depth for maxcut on large-girth regular graphs and the sherrington-kirkpatrick model},
  author={Basso, Joao and Farhi, Edward and Marwaha, Kunal and Villalonga, Benjamin and Zhou, Leo},
  journal={arXiv preprint arXiv:2110.14206},
  year={2021}
}

@article{bergholm2018pennylane,
  title={Pennylane: Automatic differentiation of hybrid quantum-classical computations},
  author={Bergholm, Ville and Izaac, Josh and Schuld, Maria and Gogolin, Christian and Ahmed, Shahnawaz and Ajith, Vishnu and Alam, M Sohaib and Alonso-Linaje, Guillermo and AkashNarayanan, B and Asadi, Ali and others},
  journal={arXiv preprint arXiv:1811.04968},
  year={2018}
}

@article{erdHos2013spectral,
  title={Spectral statistics of Erd{\H{o}}s--R{\'e}nyi graphs I: Local semicircle law},
  author={Erd{\H{o}}s, L{\'a}szl{\'o} and Knowles, Antti and Yau, Horng-Tzer and Yin, Jun},
  year={2013}
}

@article{albert2002statistical,
  title={Statistical mechanics of complex networks},
  author={Albert, R{\'e}ka and Barab{\'a}si, Albert-L{\'a}szl{\'o}},
  journal={Reviews of modern physics},
  volume={74},
  number={1},
  pages={47},
  year={2002},
  publisher={APS}
}

@article{bianconi2001bose,
  title={Bose-Einstein condensation in complex networks},
  author={Bianconi, Ginestra and Barab{\'a}si, Albert-L{\'a}szl{\'o}},
  journal={Physical review letters},
  volume={86},
  number={24},
  pages={5632},
  year={2001},
  publisher={APS}
}

@article{watts1998collective,
  title={Collective dynamics of ‘small-world’networks},
  author={Watts, Duncan J and Strogatz, Steven H},
  journal={nature},
  volume={393},
  number={6684},
  pages={440--442},
  year={1998},
  publisher={Nature Publishing Group}
}

@inproceedings{choi2019tutorial,
  title={A tutorial on quantum approximate optimization algorithm (QAOA): Fundamentals and applications},
  author={Choi, Jaeho and Kim, Joongheon},
  booktitle={2019 international conference on information and communication technology convergence (ICTC)},
  pages={138--142},
  year={2019},
  organization={IEEE}
}

@inproceedings{lin2024towards,
  title={Towards optimizations of quantum circuit simulation for solving max-cut problems with qaoa},
  author={Lin, Yu-Cheng and Wang, Chuan-Chi and Tu, Chia-Heng and Hung, Shih-Hao},
  booktitle={Proceedings of the 39th ACM/SIGAPP Symposium on Applied Computing},
  pages={1487--1494},
  year={2024}
}

@article{ai2024quantum,
  title={Quantum error correction below the surface code threshold},
  author={{Google Quantum AI}},
  journal={Nature},
  volume={638},
  number={8052},
  pages={920},
  year={2024}
}

@inproceedings{liang2023hybrid,
  title={Hybrid gate-pulse model for variational quantum algorithms},
  author={Liang, Zhiding and Song, Zhixin and Cheng, Jinglei and He, Zichang and Liu, Ji and Wang, Hanrui and Qin, Ruiyang and Wang, Yiru and Han, Song and Qian, Xuehai and others},
  booktitle={2023 60th ACM/IEEE Design Automation Conference (DAC)},
  pages={1--6},
  year={2023},
  organization={IEEE}
}

@inproceedings{chen2023quantum,
  title={Quantum reinforcement learning for quantum architecture search},
  author={Chen, Samuel Yen-Chi},
  booktitle={Proceedings of the 2023 international workshop on quantum classical cooperative},
  pages={17--20},
  year={2023}
}

@inproceedings{he2024parameter,
  title={Parameter setting heuristics make the quantum approximate optimization algorithm suitable for the early fault-tolerant era},
  author={He, Zichang and Shaydulin, Ruslan and Herman, Dylan and Li, Changhao and Raymond, Rudy and Sureshbabu, Shree Hari and Pistoia, Marco},
  booktitle={Proceedings of the 43rd IEEE/ACM International Conference on Computer-Aided Design},
  pages={1--7},
  year={2024}
}

@article{hao2024end,
  title={End-to-end protocol for high-quality QAOA parameters with few shots},
  author={Hao, Tianyi and He, Zichang and Shaydulin, Ruslan and Larson, Jeffrey and Pistoia, Marco},
  journal={arXiv preprint arXiv:2408.00557},
  year={2024}
}

@article{he2023alignment,
  title={Alignment between initial state and mixer improves QAOA performance for constrained optimization},
  author={He, Zichang and Shaydulin, Ruslan and Chakrabarti, Shouvanik and Herman, Dylan and Li, Changhao and Sun, Yue and Pistoia, Marco},
  journal={npj Quantum Information},
  volume={9},
  number={1},
  pages={121},
  year={2023},
  publisher={Nature Publishing Group UK London}
}

@article{grimsley2019adaptive,
  title={An adaptive variational algorithm for exact molecular simulations on a quantum computer},
  author={Grimsley, Harper R and Economou, Sophia E and Barnes, Edwin and Mayhall, Nicholas J},
  journal={Nature communications},
  volume={10},
  number={1},
  pages={3007},
  year={2019},
  publisher={Nature Publishing Group UK London}
}

@article{holmes2022connecting,
  title={Connecting ansatz expressibility to gradient magnitudes and barren plateaus},
  author={Holmes, Zo{\"e} and Sharma, Kunal and Cerezo, Marco and Coles, Patrick J},
  journal={PRX quantum},
  volume={3},
  number={1},
  pages={010313},
  year={2022},
  publisher={APS}
}

@article{mitarai2018quantum,
  title={Quantum circuit learning},
  author={Mitarai, Kosuke and Negoro, Makoto and Kitagawa, Masahiro and Fujii, Keisuke},
  journal={Physical Review A},
  volume={98},
  number={3},
  pages={032309},
  year={2018},
  publisher={APS}
}

@article{xue2021effects,
  title={Effects of quantum noise on quantum approximate optimization algorithm},
  author={Xue, Cheng and Chen, Zhao-Yun and Wu, Yu-Chun and Guo, Guo-Ping},
  journal={Chinese Physics Letters},
  volume={38},
  number={3},
  pages={030302},
  year={2021},
  publisher={IOP Publishing}
}

@article{wang2020xy,
  title={XY mixers: Analytical and numerical results for the quantum alternating operator ansatz},
  author={Wang, Zhihui and Rubin, Nicholas C and Dominy, Jason M and Rieffel, Eleanor G},
  journal={Physical Review A},
  volume={101},
  number={1},
  pages={012320},
  year={2020},
  publisher={APS}
}

@article{hadfield2019quantum,
  title={From the quantum approximate optimization algorithm to a quantum alternating operator ansatz},
  author={Hadfield, Stuart and Wang, Zhihui and O’gorman, Bryan and Rieffel, Eleanor G and Venturelli, Davide and Biswas, Rupak},
  journal={Algorithms},
  volume={12},
  number={2},
  pages={34},
  year={2019},
  publisher={MDPI}
}

@article{he2025non,
  title={Non-variational quantum random access optimization with alternating operator ansatz},
  author={He, Zichang and Raymond, Rudy and Shaydulin, Ruslan and Pistoia, Marco},
  journal={Scientific Reports},
  volume={15},
  number={1},
  pages={29191},
  year={2025},
  publisher={Nature Publishing Group UK London}
}

@article{vijendran2024expressive,
  title={An expressive ansatz for low-depth quantum approximate optimisation},
  author={Vijendran, V and Das, Aritra and Koh, Dax Enshan and Assad, Syed M and Lam, Ping Koy},
  journal={Quantum Science and Technology},
  volume={9},
  number={2},
  pages={025010},
  year={2024},
  publisher={IOP Publishing}
}

@inproceedings{jang2023quixote,
  title={Quixote: Improving Fidelity of Quantum Program by Independent Execution of Controlled Gates},
  author={Jang, Enhyeok and Choi, Seungwoo and Ro, Won Woo},
  booktitle={2023 60th ACM/IEEE Design Automation Conference (DAC)},
  pages={1--6},
  year={2023},
  organization={IEEE}
}

@article{jang2025balancing,
  title={Balancing Thermal Relaxation Deviations of Near-Future Quantum Computing Results via Bit-Inverted Programs},
  author={Jang, Enhyeok and Kim, Youngmin and Seo, Jeewoo and Choi, Seungwoo and Ro, Won Woo},
  journal={arXiv preprint arXiv:2502.20710},
  year={2025}
}

@article{lee2025pimutation,
  title={PIMutation: Exploring the Potential of PIM Architecture for Quantum Circuit Simulation},
  author={Lee, Dongin and Jang, Enhyeok and Choi, Seungwoo and An, Junwoong and Kim, Cheolhwan and Ro, Won Woo},
  journal={arXiv preprint arXiv:2503.00668},
  year={2025}
}

@article{kim2024distribution,
  title={Distribution-Adaptive Dynamic Shot Optimization for Variational Quantum Algorithms},
  author={Kim, Youngmin and Jang, Enhyeok and Kim, Hyungseok and Choi, Seungwoo and Lee, Changhun and Kim, Donghwi and Kyoung, Woomin and Shin, Kyujin and Ro, Won Woo},
  journal={arXiv preprint arXiv:2412.17485},
  year={2024}
}

@inproceedings{kim2025qr,
  title={QR-Map: A Map-Based Approach to Quantum Circuit Abstraction for Qubit Reuse Optimization},
  author={Kim, Hyungseok and Jang, Enhyeok and Choi, Seungwoo and Kim, Youngmin and Ro, Won Woo},
  booktitle={Proceedings of the 52nd Annual International Symposium on Computer Architecture},
  pages={1568--1582},
  year={2025}
}

@article{jang2026plutarch,
  title={Plutarch: Toward Scalable Operational Parallelism on Racetrack-Shaped Trapped-Ion Processors},
  author={Jang, Enhyeok and Kim, Hyungseok and Lee, Yongju and Kwon, Jaewon and Huang, Yipeng and Ro, Won Woo},
  journal={arXiv preprint arXiv:2601.08930},
  year={2026}
}

\end{document}